\documentclass[journal=nalefd,manuscript=letter]{achemso}

\usepackage{siunitx}
\usepackage[version=4]{mhchem}

\author{Marinus C. van der Maas}
\email{m.c.vandermaas@tudelft.nl}
\affiliation[Delft University of Technology]{Quantum and Computer Engineering Department, Delft University of Technology, Delft, Netherlands}
\alsoaffiliation[QuTech]{QuTech, Delft University of Technology, Delft}
\author{Lin Jin}
\email{l.jin@tudelft.nl}
\affiliation[Delft University of Technology]{Quantum and Computer Engineering Department, Delft University of Technology, Delft, Netherlands}
\alsoaffiliation[QuTech]{QuTech, Delft University of Technology, Delft}
\alsoaffiliation[Kavli]{Kavli Institute of Nanoscience, Delft University of Technology, Delft}
\author{Ilhan Tunç}
\affiliation[Delft University of Technology]{Quantum and Computer Engineering Department, Delft University of Technology, Delft, Netherlands}
\author{Raymond Vermeulen}
\author{Henri~Ervasti}
\affiliation[QuTech]{QuTech, Delft University of Technology, Delft}
\author{Ravi Gopie}
\affiliation[Delft University of Technology]{Quantum and Computer Engineering Department, Delft University of Technology, Delft, Netherlands}
\author{Jan Riegelmeyer}
\affiliation[Delft University of Technology]{Quantum and Computer Engineering Department, Delft University of Technology, Delft, Netherlands}
\alsoaffiliation[QuTech]{QuTech, Delft University of Technology, Delft}
\alsoaffiliation[Kavli]{Kavli Institute of Nanoscience, Delft University of Technology, Delft}
\author{Marco~Colangelo}
\affiliation[Northeastern]{Department of Electrical and Computer Engineering, Northeastern University, Boston, USA}
\author{Ryoichi~Ishihara}
\affiliation[Delft University of Technology]{Quantum and Computer Engineering Department, Delft University of Technology, Delft, Netherlands}
\alsoaffiliation[QuTech]{QuTech, Delft University of Technology, Delft}
\author{Carlos~Errando-Herranz}
\email{c.errandoherranz@tudelft.nl}
\affiliation[Delft University of Technology]{Quantum and Computer Engineering Department, Delft University of Technology, Delft, Netherlands}
\alsoaffiliation[QuTech]{QuTech, Delft University of Technology, Delft}
\alsoaffiliation[Kavli]{Kavli Institute of Nanoscience, Delft University of Technology, Delft}

\title[An \textsf{achemso} demo]
  {Geometry-enabled magnetic resilience in superconducting nanowire single-photon detectors}

\abbreviations{SNSPD,CR}

\begin{document}

\begin{abstract} 
While magnetic fields and superconductors are both central to classical and quantum technologies, their combined use is often challenging, as magnetic fields significantly affect superconducting device performance.
In superconducting nanowire single-photon detectors (SNSPDs), magnetic fields drastically reduce detection efficiencies, hampering their application in magnetically-active classical and quantum photonics.
Here, we systematically characterize the performance of NbTiN SNSPDs under magnetic fields and show the enhancement of their intrinsic detection efficiency (IDE) at lower bias currents and its suppression at higher currents.
This leads to SNSPD performance degradation through reduced or disappearing saturation plateaus.
We show that the magnitude of this degradation is highly dependent on nanowire width and demonstrate width-optimized SNSPDs with saturating IDE for a wide range of photon energies under application-relevant magnetic fields. 
Minimizing degradation in superconducting devices under magnetic fields enables applications like detector-integrated spin-optic and atomic quantum processors, high-sensitivity magnetometry, and quantum transduction.

\end{abstract}

\section{Introduction}

Magnetic fields are central to qubit formation, control, and tuning in spin\cite{awschalom2018quantum}, atomic\cite{ebadi2021quantum,bruzewicz2019trapped}, and superconducting platforms\cite{kjaergaard2020superconducting}, as well as to a range of precision experiments\cite{koppell2025dark, armstrong2015threshold, polakovic_superconducting_2020}. 
Superconducting films, meanwhile, underpin leading quantum processors\cite{google2025quantum} and enable high-performance photon detection via superconducting nanowire single-photon detectors (SNSPDs)\cite{venza2025research}.
SNSPDs offer high efficiency\cite{reddy2020superconducting}, detection of low energy photons \cite{verma2021single}, excellent timing resolution \cite{korzh2020demonstration}, and compatibility with integrated photonics \cite{ferrari_waveguide-integrated_2018}, yet their operation in magnetically active environments remains challenging. 
This challenge is particularly relevant in platforms such as optically controlled atomic systems\cite{degen2017quantum, ebadi2021quantum, bruzewicz2019trapped}, spin-photon magnetometers\cite{barry2020sensitivity}, dark matter detection~\cite{koppell2025dark}, and nuclear physics experiments~\cite{armstrong2015threshold, polakovic_superconducting_2020}, where strong magnetic fields and superconducting detection must coexist.

Superconducting nanowires are intrinsically sensitive to magnetic fields, which modify their critical current and detection characteristics.
Magnetic fields modulate the critical current density in nanowires, contributing to an already complex interplay between geometrical and bend parameters \cite{hortensius2012critical, henrich2012geometry, semenov_asymmetry_2015}, edge defects \cite{kerman2007constriction, chen2023visualizing, charaev2017enhancement}, inhomogeneities \cite{liu2025revealing}, cooling efficiency \cite{zhang2018hotspot}, the presence of phase-slip centers \cite{klimov_characterization_2017, ejrnaes2019superconductor}, the depairing current density\cite{ilin_critical_2005, ilin_influence_2010}, and vortex entry dynamics \cite{bulaevskii2011vortex}. 
Through the latter effect, magnetic fields affect the intrinsic detection efficiency (IDE) of SNSPDs. 
Magnetic-field-induced screening currents enhance IDE at lower bias currents and suppress it at higher bias currents, leading to what we refer to as the IDE enhancement-suppression transition (EST) effect.
Combined with the reduction in critical current, the EST results in a strongly reduced or absent saturation plateau and thus below-unity IDE.
Previous work has suggested a relation between the EST effect magnitude, the hotspot size and thus the photon energy\cite{vodolazov_vortex-assisted_2015}, and the nanowire width\cite{korneeva2020different}.
While these studies suggest a relation between detection efficiency and SNSPD geometry, no comprehensive study exists about the effect of nanowire geometry on SNSPD performance in magnetic fields, and thus, no guidelines exist for the design of magnetically-resilient SNSPDs.

Here, we perform a systematic experimental study of the effect of SNSPD geometry on their detection performance under \qty{130}{~mT} applied magnetic fields at three different wavelengths.
We observe that the magnitude of the EST effect and in turn the degradation of the saturation plateau strongly depends on nanowire width, and demonstrate SNSPDs with unity IDE over multiple wavelengths under application-relevant magnetic fields.

\section{Methods}

\begin{figure}
    \includegraphics[width=1\textwidth]{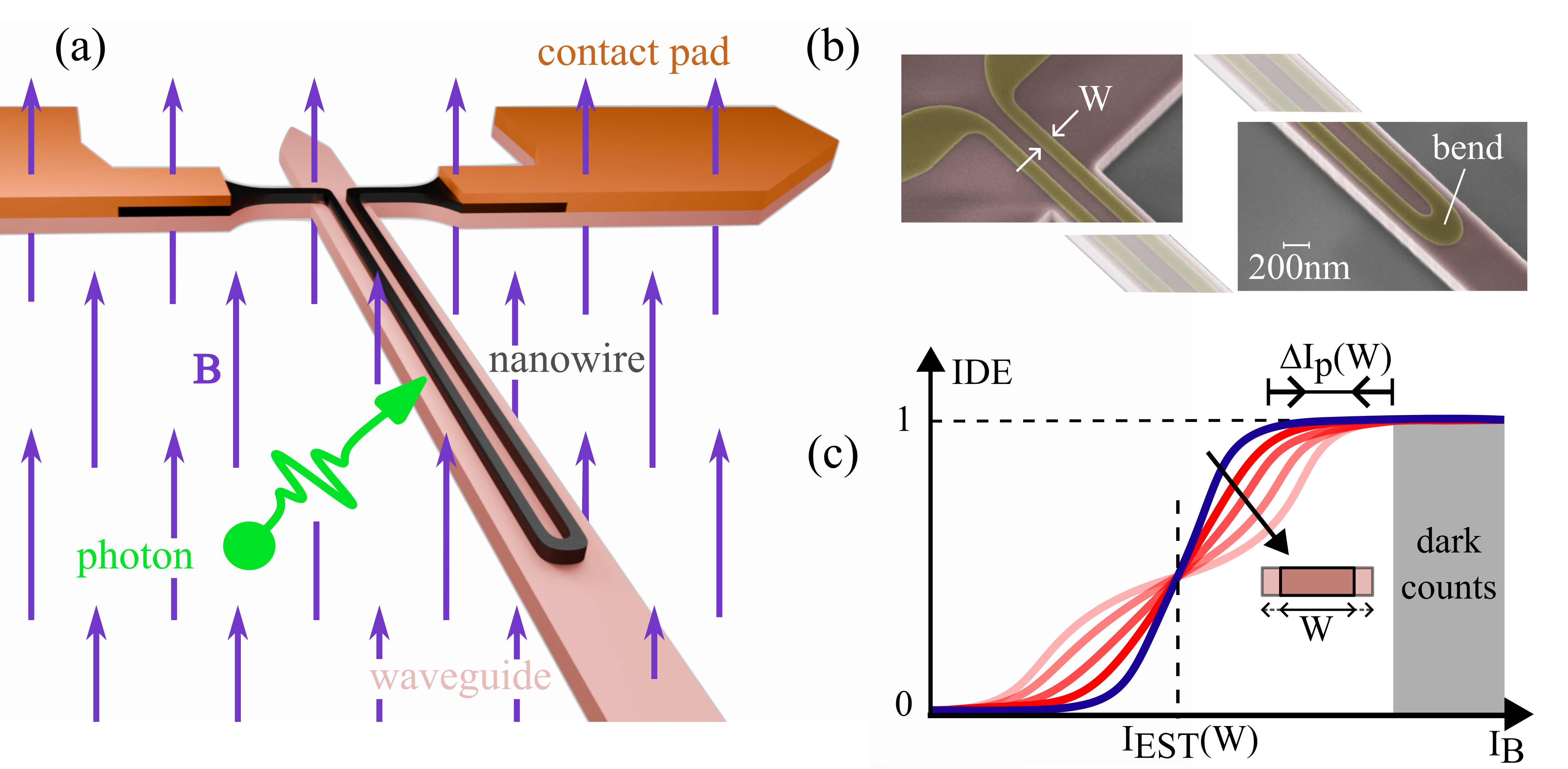}
    \caption{(a) Schematic of a waveguide-integrated SNSPD under an applied magnetic field and optical flood illumination. (b) Scanning electron microscope (SEM) images of one of our devices. 
    (c) Schematic showing the effect of magnetic field on the IDE of SNSPDs for different widths W: enhancement for lower $I_B$ and suppression for higher $I_B$ leading to a reduction of the near-unity IDE plateau $\Delta I_p(W)$.}
    \label{fig:concept_fig}
\end{figure}

Our devices consist of \ce{NbTiN} nanowires on top of \ce{Si_3 N_4} waveguides, as shown in Fig.~\ref{fig:concept_fig}(a). 
The \ce{NbTiN} film thickness ranges from \qty{6}{nm} to \qty{10}{nm} and widths range from \qty{65}{nm} to \qty{135}{nm}. 
Details about the device fabrication can be found in the Supplementary Information (SI). 
Our nanowires feature a single hairpin bend with an optimized curvature that mitigates current crowding\cite{clem_geometry-dependent_2011, jonsson_current_2022}.
This single-bend also minimizes the effect of additional magnetically-induced current crowding due to its uniform bend direction \cite{semenov_asymmetry_2015, henrich2013detection, charaev_magnetic-field_2019}.
Figure~\ref{fig:concept_fig}(b) shows false-colored electron microscopy images of one of the SNSPDs. 
The SNSPDs were characterized at a temperature of \qty{~1}{K}.
Laser light was fiber and free-space coupled into the cryostat, and the resulting detection pulses were amplified at room temperature with low-noise amplifiers (LNAs) by approximately \qty{46}{dB}. 
The counts of multiple SNSPDs were registered simultaneously with a time tagger. 
All measurements were performed both with and without a magnetic field of $130\pm$\qty{23}{mT} applied perpendicular to the sample substrate, leading to width-dependent effects as schematically shown in Fig~\ref{fig:concept_fig} (c). 
More details about the measurement setup and the applied magnetic field can be found in the SI.

\section{Results}

To understand how the superconducting nanowire dimensions affect the current density profile, we consider a nanowire of thickness $t$ and width $W$. 
We focus on the regime $W\ll\Lambda$. 
Here, the penetration depth $\lambda$ characterizes the decay length of magnetic fields into the bulk of a superconductor, whereas the Pearl length $\Lambda=2\lambda^2/t$ represents the characteristic in-plane magnetic-screening length of a thin superconducting film ($t\ll\lambda$) \cite{pearl1964current}.
A magnetic field $B_{\perp}$ is applied perpendicular to the film plane. 
The resulting screening current distribution can be described in terms of the sheet current density $j_s$ along the coordinate $x$ of the nanowire cross-section. 
For nanowire widths in the regime $W\ll\Lambda$, self-screening effects are negligible, so the magnetic field can be treated as uniform across the width. 
In the Meissner state (i.e., in the absence of vortex penetration) $j_s$ then satisfies \cite{gaggioli2024superconductivity}
\begin{equation}
    \frac{dj_s}{dx}\approx\frac{c}{2\pi\Lambda}B_{\perp},
\end{equation}
where $c$ is the speed of light.
Therefore, the difference in bias current density between the two edges along the nanowire is 
\begin{equation}
    \Delta j_s = \frac{Wc}{2\pi\Lambda}B_{\perp}.
    \label{eq:DeltaJs}
\end{equation}
This indicates that $\Delta j_s$ scales linearly both with $W$ and $t$, lowering the vortex entry barrier along one nanowire edge and increasing it along the other.

\begin{figure}
    \includegraphics[width=0.5\textwidth]{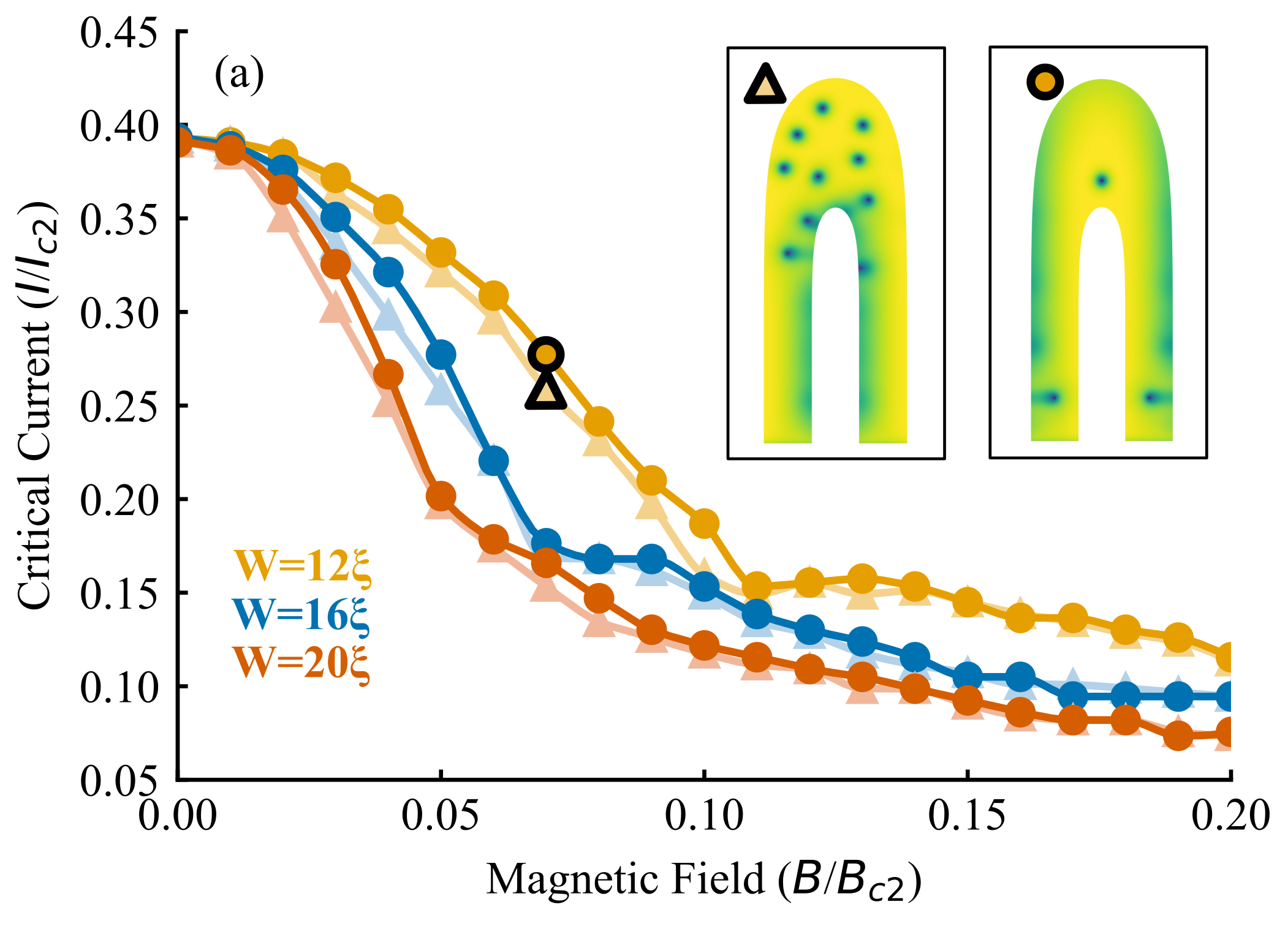}%
    \includegraphics[width=0.5\textwidth]{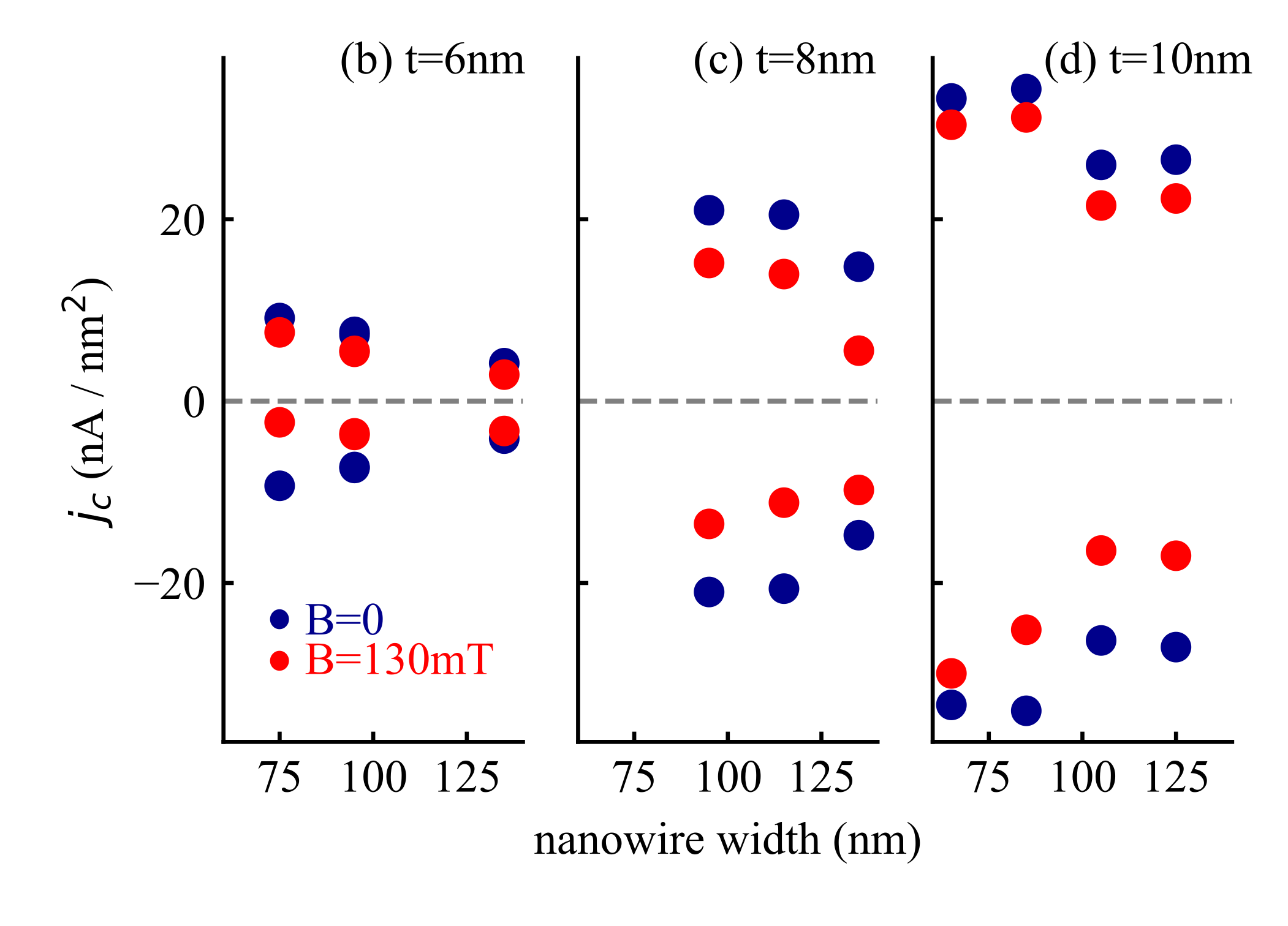}%
    \caption{(a) Simulated $I_c$ as a function of B for different nanowire widths. Circular/triangular markers correspond to positive/negative bias conditions. (b-d) Measured $j_c$ as a function of nanowire width for film thickness of (b) \qty{6}{nm}, (c) \qty{8}{nm} and (d) \qty{10}{nm}, with and without magnetic field.}
    \label{fig:Ic_reduc}
\end{figure}

This effect can be confirmed via simulations.
Fig.~\ref{fig:Ic_reduc}(a) shows a Ginzburg-Landau (GL) simulation of the critical current $I_c$ as a function of the magnetic field for several nanowire widths. 
We adapted the simulation setup (SuperDetectorPy) from \citet{jonsson_current_2022} \cite{jonsson2022theory} to simulate the magnetic field resilience of the hairpin bend geometry of our SNSPDs. 
Nanowire widths are chosen to be \qty{12}{\xi}, \qty{16}{\xi} and \qty{20}{\xi}, corresponding respectively to a width of \qty{54}{nm}, \qty{72}{nm} and \qty{90}{nm}, assuming a coherence length $\xi=4.5$~nm\cite{mironov2018charge}. 
The simulation results suggest that the reduction in $I_c$ is less pronounced when the nanowire is narrow, similar to Eq.~\ref{eq:DeltaJs}.
We performed simulations with both positive and negative perpendicular magnetic fields, which is equivalent to reversing the bias current direction. 
The simulation results in Fig.~\ref{fig:Ic_reduc}(a) predict that the critical current depends on the magnetic field (and thus bias current) direction. 
The insets in Fig.~\ref{fig:Ic_reduc}(a) show vortex-crossing when the bias current is equal to the critical current. 
Under positive bias conditions, vortices nucleate in the straight wire sections, while under negative bias conditions, vortices nucleate from the inside of the bends, limiting $I_c$.

We measured the critical current density $j_c$ for all devices, shown in Fig.~\ref{fig:Ic_reduc}(b), (c) and (d) for nanowires with thickness \qty{6}{nm}, \qty{8}{nm}, and \qty{10}{nm} respectively. 
We observed that narrower nanowires reach a higher $j_c$. 
Simultaneously, the thicker samples switch at a higher bias currents density. 
Our $j_c$ measurements in Fig.~\ref{fig:Ic_reduc}(b), (c), (d) confirm that the critical current does depend on bias current direction when a magnetic field is applied. 
The magnitude of this bias polarity dependence, however, is more significant than what the simulations predict. This is likely due to the geometrically not optimized outgoing bends of our SNSPDs (see SI). 

\begin{figure}
    \includegraphics[width=1\textwidth]{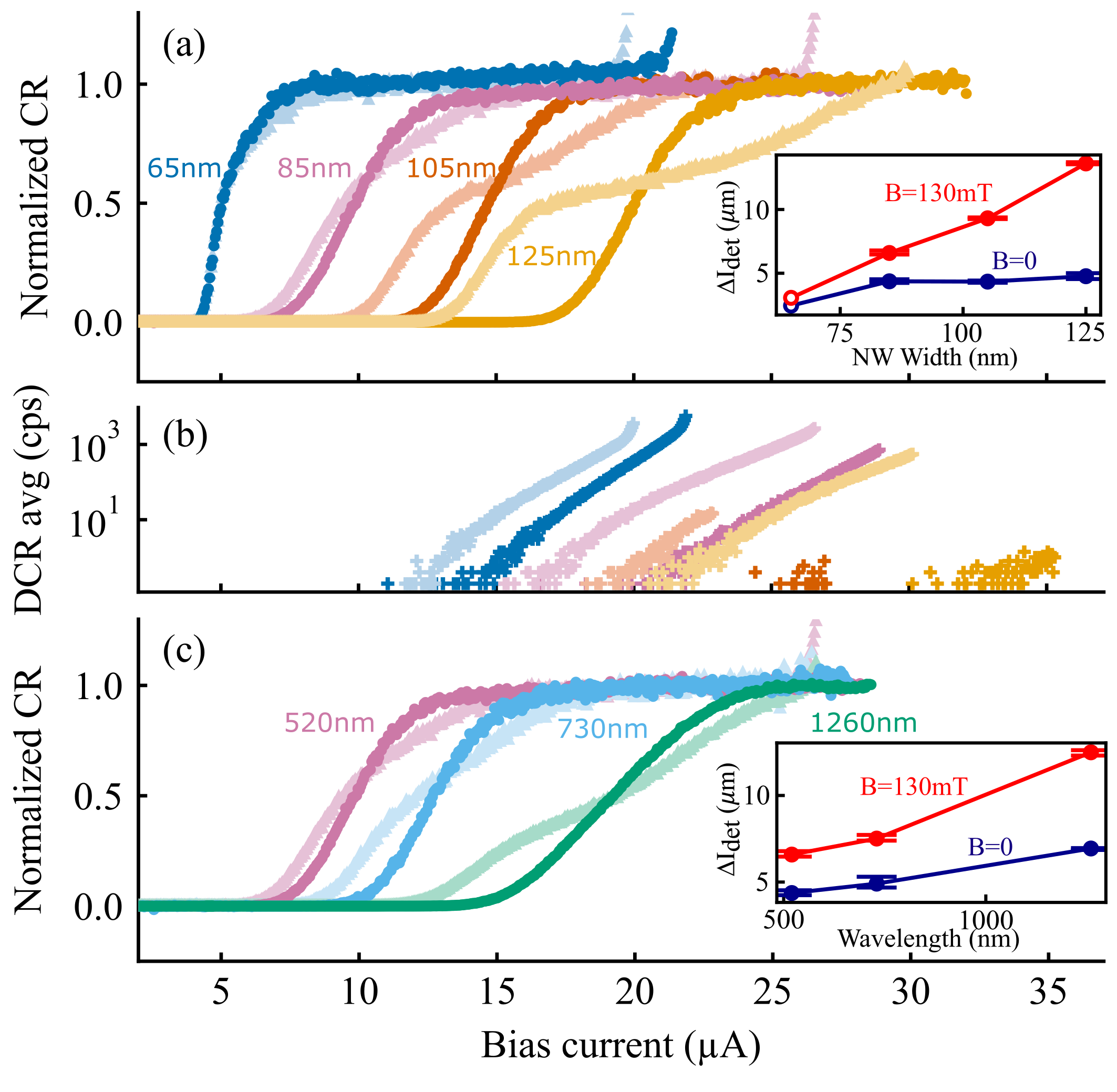}%
    \caption{(a) PCR of \qty{10}{nm}-thick SNSPDs with different nanowire widths, measured with and without magnetic field. The dark circles correspond to $B=0$ and the lighter triangles correspond to $B=\qty{130}{mT}$. (b) Dark counts rates. (c) Measurement results for the $\qty{85}{nm}$-wide sample shown in (a), under illumination at different wavelengths. The corresponding DCR is also shown in purple in (b). The insets shows $\Delta I_\text{det}$.}
    \label{fig:IDE_width_measurement}
\end{figure}

Following the static analysis, we evaluated the SNSPD detection performance for different widths, with and without magnetic fields. 
See SI for details about the acquisition methods.
Fig.~\ref{fig:IDE_width_measurement}(a) and (b) illustrate the normalized photon count rate (PCR) and corresponding dark count rate (DCR) for nanowires of varying widths under \qty{520}{nm} illumination.
To help in our analysis, we define $I_\text{det,min}$ and $I_\text{det,max}$ as the bias currents at which the IDE of the SNSPD reaches 10\% and 90\% respectively. 
Here, $I_\text{det}$ denotes the position-dependent threshold current, above which a photon absorption event initiates the detection mechanism in the nanowire \cite{zotova_intrinsic_2014}. 

As expected, for zero external magnetic field, the PCR measurement follows a sigmoid curve, with IDE saturating near unity for higher bias currents. 
In the absence of external magnetic field, increasing the nanowire width shifts the IDE curve toward higher bias currents, thereby increasing both $I_\text{det,min}$ and $I_\text{det,max}$, and leaving $\Delta I_\text{det} = I_\text{det,max} - I_\text{det,min}$ nearly unchanged, as shown in the inset of Figure~\ref{fig:IDE_width_measurement}(a). 
This is consistent with the hotspot model framework\cite{semenov2001quantum, natarajan2012superconducting}, in which the photon-induced normal region occupies a smaller fraction of the cross-section in wider nanowires, resulting in weaker current crowding and higher bias current to trigger a resistive transition. 
Note that due to the counter discriminator voltage, which is needed for filtering out noise peaks, detection events below a certain bias currents cannot be registered. This likely leads to an overestimated $I_\text{det,min}$ value and underestimated $\Delta I_\text{det}$ value for the \qty{65}{nm} wide nanowire.

Under an applied magnetic field, we observed a reduction in $I_\text{det,min}$ and an increase in $I_\text{det,max}$.
This EST effect likely arises from the magnetic-field-induced redistribution of current density across the nanowire width. 
The effect is apparent for wider nanowires, which show an significantly increased $\Delta I_\text{det}$.
In contrast, the narrowest nanowires exhibit a nearly unaffected performance.
This can be attributed to the nanowire width-dependent redistribution of the screening current density $\Delta j_s$ (see Eq. \ref{eq:DeltaJs}). 
The redistribution lowers the bias current required to reach $I_\text{det}$ near one nanowire edge, while simultaneously increasing the bias current needed for the detection condition to be satisfied across the entire nanowire width.
A detailed graphical representation of the current density profile is provided in the SI. 
Further measurement data for nanowires of varying widths and thickness of \qty{6}{nm} and \qty{8}{nm} are presented in the SI. 

Figure~\ref{fig:IDE_width_measurement}(b) depicts the DCR of the detectors as a function of the bias currents for different nanowire widths. 
Dark counts are generally modeled as arising from thermally activated processes that enable vortex entry and crossing of the nanowire. \cite{bulaevskii_vortex-assisted_2012}. 
Experimentally, the measured DCR can be influenced by factors like localized film defects \cite{zhang_characterization_2014} and current crowding at bends \cite{clem_geometry-dependent_2011}, both of which can locally reduce the vortex-entry barrier.  
We observe that applying a perpendicular magnetic field increases the DCR at a fixed bias current. 
This behavior can be attributed to a reduction of the energy barrier for vortex entry \cite{vodolazov_vortex-assisted_2015}, which enhances the rate of thermally activated vortex crossings. 
The same mechanism also suppresses the critical current, consistent with the decrease in critical current density observed in Fig.~2(a). 
Accordingly, the shift in the DCR characteristics follows the corresponding reduction in critical current.

We also characterized the nanowires under illumination at various wavelengths, as shown in Fig.~\ref{fig:IDE_width_measurement}(c). 
For no magnetic field, as expected, the onset of detection events shifts toward higher bias currents as the wavelength increases. 
Additionally, the EST effect becomes more pronounced at longer wavelengths in the presence of magnetic field, as shown in Fig.~\ref{fig:IDE_width_measurement}(c) inset. 
These observations can be attributed to longer wavelengths photons carrying lower energies and thus generating smaller hot-spots \cite{semenov2001quantum, vodolazov_single-photon_2017}. 
This is phenomenologically similar to increasing the nanowire width at a constant wavelength, and thus consistent with our results in Fig~\ref{fig:IDE_width_measurement}(a). 

\begin{figure}
    \includegraphics[width=1\textwidth]{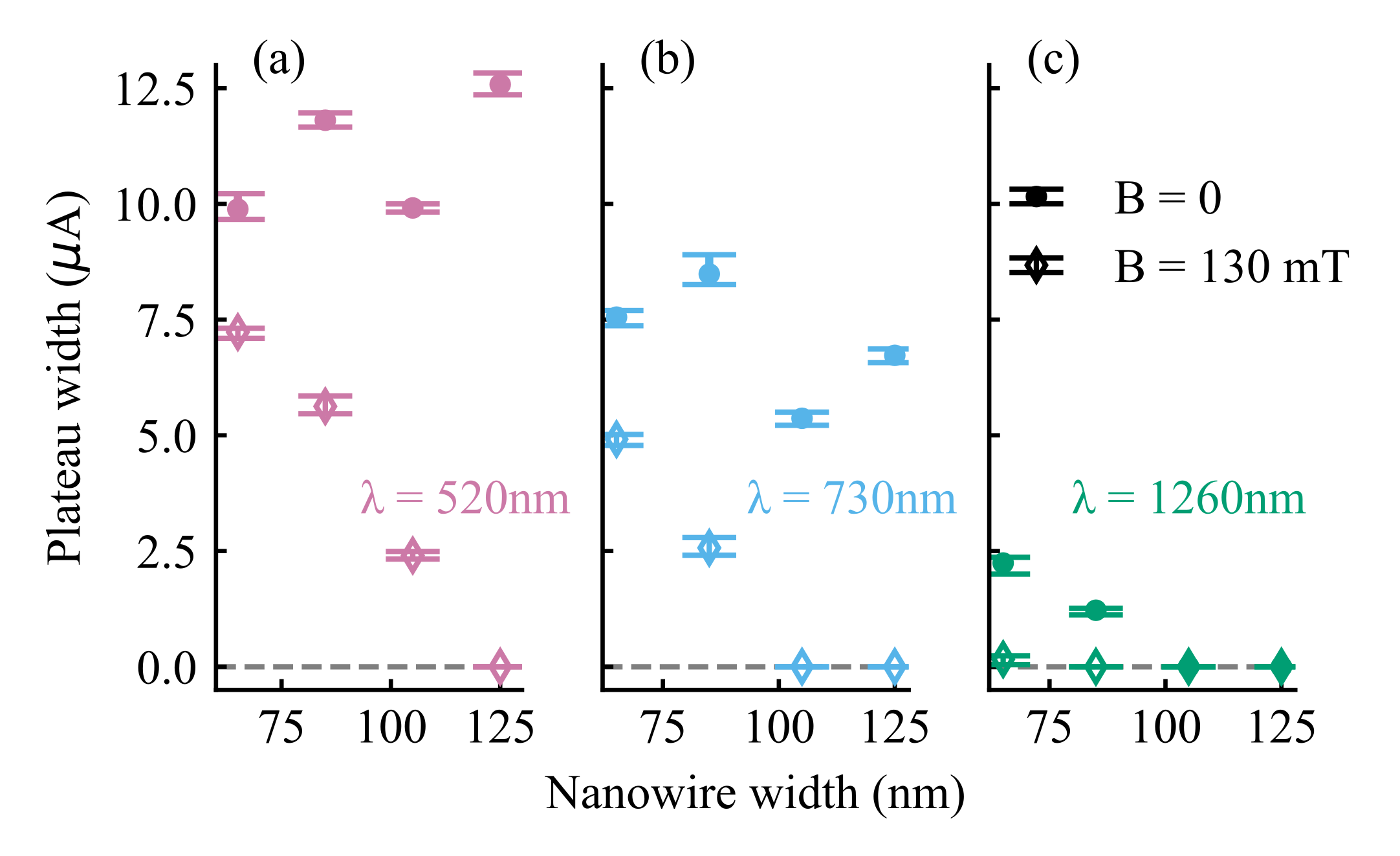}%
    \caption{Plateau width of the \qty{10}{nm} thick devices for different wavelength. Open and filled markers represent the performance with and without magnetic field, respectively. The data shows that narrower nanowires approach the performance of zero-field nanowires under a magnetic field.}
    \label{fig:FOM_10nm}
\end{figure}

A critical value for SNSPDs is the range of bias currents at which IDE saturates to near-unity, i.e. the IDE plateau width.
Wider plateaus are better for practical applications, since the operation bias current is often chosen as the one achieving highest IDE for a DCR below the acceptable threshold. 
Figure~\ref{fig:FOM_10nm} shows the width of the IDE saturation plateau for the \qty{10}{nm} thick devices, defined as the range from $I_\text{det,max}$ to the bias current at which the DCR reaches~\qty{10}{cps}. 
In the absence of a magnetic field, we observed no significant detector width dependence of the plateau width.
Small variations in the detection performance for the same wavelength as a function of width are likely dominated by device fabrication quality.
Conversely, under the magnetic field, the plateau width decreases with increasing nanowire width across a broadband wavelength range, illustrating the influence of magnetic field and device geometry. 
Given that narrower nanowires exhibit broader detection plateau, extrapolating the trends shown in Fig.~\ref{fig:FOM_10nm} suggests that narrower devices could reach negligible degradation in the applied magnetic field.

\section{Discussion}

Our experiments bring a comprehensive set of data to the limited literature reports on magnetic field effects in superconducting nanowires.
Our observed EST and $\Delta I_\text{det}$ scaling with the nanowire width $W$ and the incident light wavelength $\lambda$ under magnetic fields are in line with some of the earlier research.
For example, Ref.~\cite{vodolazov_vortex-assisted_2015} experimentally shows a similar EST for a \qty{102}{nm} wide NbN SNSPD in a perpendicular magnetic field, where EST becomes more pronounced at longer wavelengths. 
In addition, the \ce{Mo_{x} Si_{1-x}} material system reported observations both for IDE transition shifts without EST~\cite{korneev_characterization_2015} for 100~nm scale wires, and EST for micron-scale wide nanowires~\cite{korneeva2020different}. 
Both our results and the aforementioned studies suggest that the occurrence of an EST and the corresponding increase of $\Delta I_\text{det}$ depend on the size of the hotspot $R$ with respect to the width of the nanowire $\Delta I_\text{det} \propto W/R$. 
However, not all studies have observed an EST. 
Ref.~\cite{renema_effect_2015, polakovic_superconducting_2020} observed only a reduction in IDE with increasing field strength for \ce{NbN} SNSPD meanders and strips with geometric dimensions comparable to those in our study. 
The discrepancies between these findings and our results may be attributed to variations in photon energy, nanowire geometry, and higher magnetic field strengths employed in those studies. 
Besides, logarithmic plotting of PCR in those studies could also hamper the identification of an EST.  

Our findings could be explained within a vortex-assisted detection framework~\cite{vodolazov_vortex-assisted_2015}, in which photon absorption locally suppresses superconductivity and the edge barrier for vortex entry gets lowered. 
Once a vortex enters the nanowire, its traversal across the width is dissipative and triggers a measurable voltage pulse\cite{jahani_probabilistic_2020}.
Redistribution of bias current due to the presence of screening currents~\cite{gaggioli2024superconductivity} lowers the barrier on one nanowire edge far more than on the other side of the cross-section~\cite{bulaevskii_vortex-assisted_2012}.
Hence, such redistribution can account for the asymmetry of the critical current with magnetic field polarity, while its interplay with vortex entry can explain the observed EST effect.

However, to validate these hypotheses and to explain the differences found between geometries and materials in literature, further work is required. 
The detection current dependence on the coordinate along the nanowire width is determined by the detection mechanism. 
The shape of this dependence is modified by the magnetic field, as predicted previously in simulations \cite{vodolazov_vortex-assisted_2015, vodolazov_single-photon_2017}. 
Further simulations over a wider range of nanowire geometries and material parameters might reveal how the dependence of $I_\text{det}$ on the absorption location, and consequently the dependence of the IDE on bias current is affected by perpendicular magnetic fields.

Experimentally, spatially selective illumination \cite{hadfield2007submicrometer} of the nanowire, or differential polarization measurements in combination with quantum detector tomography (QDT) \cite{renema_position-dependent_2015} could directly probe the effect of magnetic fields on the position-dependent detection current.
Future work may also focus on replicating our geometry-dependent observations in other popular superconductors such as \ce{Al}, \ce{WSi}, \ce{NbN} and \ce{MoSi}. 
This may include exploring narrower nanowires to potentially reach negligible magnetic-field-induced degradation, as well as wider nanowires to access a more pronounced field-dependent regime. 
Extending our study to higher fields or wider wires may also enable the exploration of regimes where vortices penetrate the wire and critical current is limited by vortex motion \cite{ceccarelli2019imaging}.

Our magnetically-resilient superconducting nanowires may readily find applications in multiple fields.
For instance, magnetic fields are required to tune energy levels and enable control of optically-detectable spin systems\cite{degen2017quantum} as in color centers and rare-earth ions, and in atomic systems such as trapped ions\cite{bruzewicz2019trapped} and neutral atoms\cite{ebadi2021quantum}.
In these applications, the magnetic fields are in the order of 100~mT as in our experiments, and the high efficiency detection of on-chip SNSPDs is required for qubit initialization, measurement, and quantum gates\cite{awschalom2018quantum}.
Another example is quantum magnetometry with spin-photon interfaces such as the nitrogen vacancies.
These systems exhibit sensitivities that scale with the square root of the detection efficiency and thus would benefit from SNSPDs integrated near the magnetically active sensor site \cite{barry2020sensitivity}.
Significant magnetic fields are also present in other sensing applications such as SNSPD-based dark matter detection \cite{koppell2025dark}, or SNSPD-based detectors in nuclear physics\cite{armstrong2015threshold, polakovic_superconducting_2020}.
The magnetic-field resilience of superconducting nanowires presented here may also benefit applications in which the nanowire is used for routing signals rather than photon detection.  
An example is quantum transduction between superconducting qubits and photons via spin-photon interfaces, where a magnetic field is required for splitting spin states while superconducting circuitry provides coupling to qubits \cite{xie2025scalable}. 
As such devices can be operated well below the field-reduced $I_c$, our work highlights the trade-off between choosing narrower wires to maximize $j_c$ and wider wires to minimize photon sensitivity.

In conclusion, we systematically studied the influence of a perpendicular magnetic field applied to hairpin bend SNSPDs with varying nanowire dimensions. 
We also showed how the nanowire width modulates the IDE through EST effect, and specifically demonstrated unity IDE for narrow-width SNSPDs over a broad biasing current and wavelength ranges.
Our results indicate that further optimization of nanowire dimensions could enable saturation plateaus at telecommunication wavelengths under stronger perpendicular magnetic fields.
These findings elucidate the role of nanowire geometry in the interaction between magnetic fields and superconductors, bringing system-level experiments and applications within reach.

\begin{acknowledgement}
The authors thank Roy Birnholtz for measurement setup support.
We gratefully acknowledge support from the joint research program “Modular quantum computers” by Fujitsu Limited and Delft University of Technology, co-funded by the Netherlands Enterprise Agency under project number PPS2007.
J.R. and C.E.-H. acknowledge funding from the Dutch Research Council (NWO, Project No. 601.QT.001).
L. J. acknowledge funding from the Netherlands Organization for Scientific Research (NWO/OCW), as part of Quantum Limits (Project No. SUMMIT.1.1016).
\end{acknowledgement}

\begin{suppinfo}

\section{Experimental setup}

\subsection{Sample preparation}

The fabrication started with wafers consisting of a \qty{250}{nm} thin stoichiometric \ce{Si_3 N_4} on top of \qty{2}{\micro\meter} \ce{Si O_2} on a silicon substrate. 
The wafers were purchased from a foundry (Rogue Valley Microdevices).
Subsequently, a thin film of \ce{Nb_x Ti_{1-x} N} of varying thickness (6, 8, and 10~nm) was commercially deposited via sputter deposition (Single Quantum).
After dicing, the SNSPDs were pattered using electron-beam lithography followed by reactive ion etching in \ce{SF_6} chemistry.
The \ce{Si_3 N_4} waveguides were then patterned via aligned electron-beam lithography followed by reactive ion etching in \ce{CHF_3} chemistry.

\subsection{Electrical setup}

Figure \ref{fig:setup_schematic} shows a schematic overview of the electrical part of the measurement setup. 

\begin{figure}
    \includegraphics[width=0.9\textwidth]{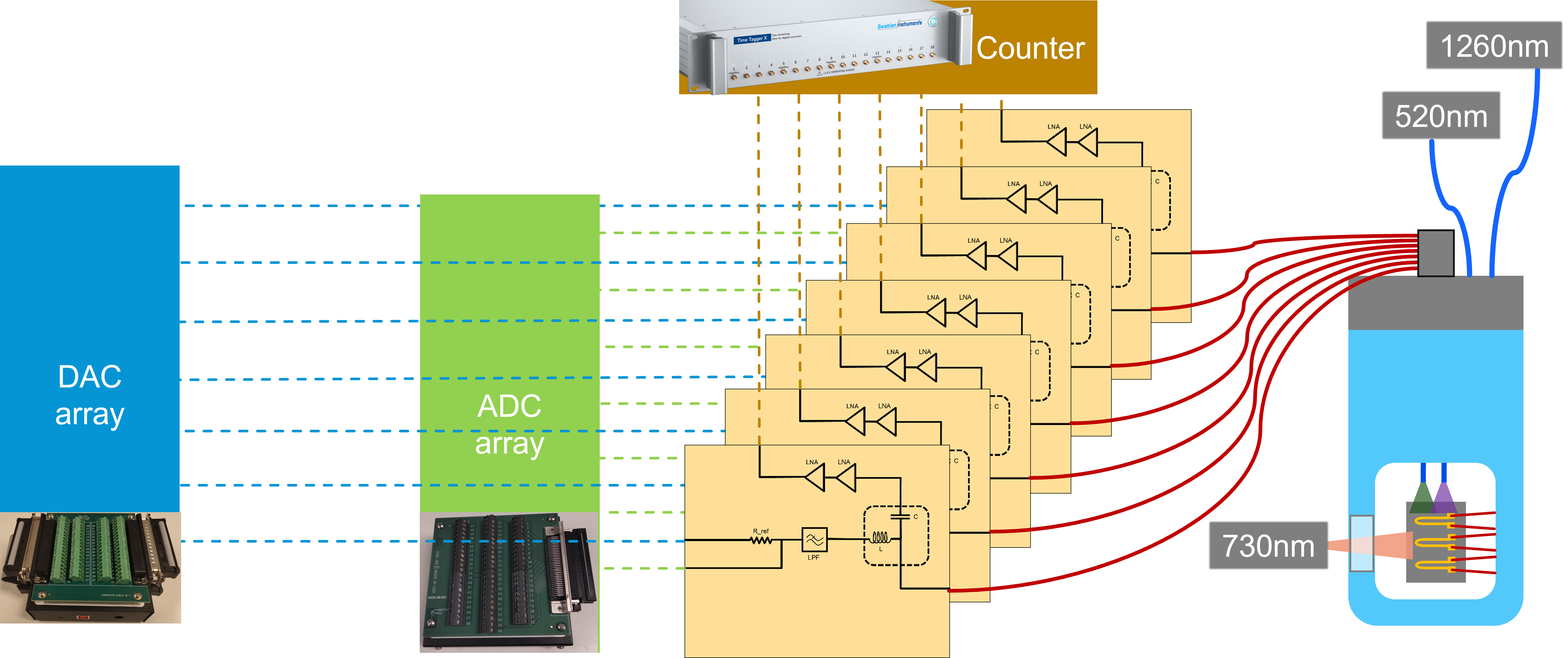}%
    \caption{Schematic overview of the measurement setup.}
    
    \label{fig:setup_schematic}
\end{figure}

The current was supplied by a digital-to-analog converter (DAC).
The voltage drop over the SNSPD and its coax connection was measured with an analog-to-digital converter (ADC).
The signal was amplified by a cascade of two Mini-Circuits ZFL-1000LN+ low noise amplifiers (LNAs). 
These amplifiers were operated with a supply voltage of \qty{15}{V}, where the gain was about \qty{23.4}{dB} and the noise figure \qty{3.09}{dB}. 
Since there were two amplifiers, the total gain was \qty{46.8}{dB} and the total noise figure $NF_T=$\qty{3.11}{dB}.

The counts were detected by a Swabian counter.

The bias tee was custom designed by the QuTech electrical engineering department. 
A simplified schematic is shown in Fig.~\ref{fig:bias_tee}. 
The current is supplied via the 'Volt Source' port, which goes to the DAC output. 
The series resistance converts the bias voltage to a a bias current, which is filtered by a low-pass filter (LPF). 
The 'Volt Meter' port measures the DC voltage over the SNSPD directly, with a filter circuit that protects the SNSPD from kickback from the ADCs. The 'AMP' port passes the detection pulses to the amplifiers.

\begin{figure}
    \includegraphics[width=0.9\textwidth]{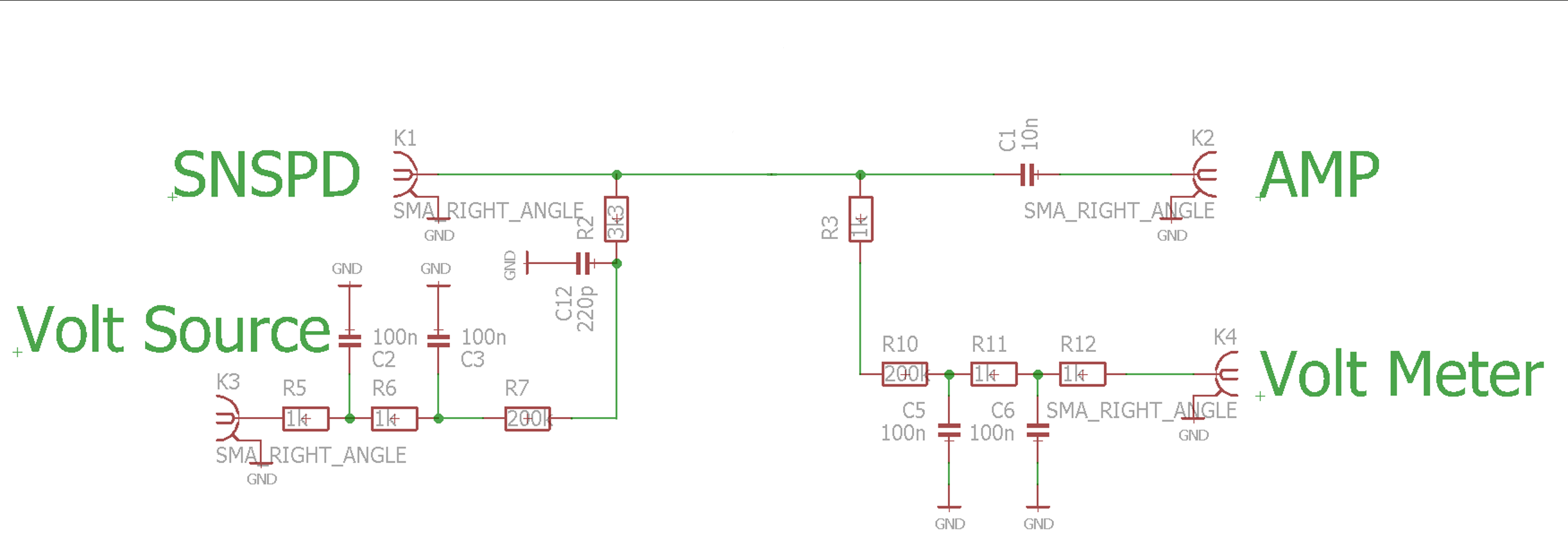}%
    \caption{Bias tee schematic.}
    
    \label{fig:bias_tee}
\end{figure}

\subsection{Magnet}

We used a neodymium (NdFeB) permanent magnet attached to the backside of the sample holder. 
Measurements at room temperature indicated a magnetic field strength of \qty{150}{mT} at the chip surface, with the field direction along the magnet axis perpendicular to the surface.
Accounting for a \qty{14}{\%} decrease in field strength due to spin reorientation at low temperatures \cite{tokuhara1985magnetization, givord1984magnetic}, we expect the field strength during measurements to be \qty{130}{mT}. 
The spatial variation causes the perpendicular component of the magnetic field to vary at most by \qty{23}{mT}, depending on the location of the SNSPD with respect to the center of the chip.

\subsection{PCR measurements}

\qty{1260}{nm} (Santec TSL-570) laser light and optical pulses at \qty{520}{nm} (Cobolt 06-MLD) were delivered via optical fiber to the detectors situated at the cryogenic stage. \qty{730}{nm} light was delivered via a view port using a Thorlabs M730L5 LED.
The resulting SNSPD pulses under a sweep of bias current were registered and recorded as the total count rate. 
To isolate the device response to photons, the dark count rate (DCR) was measured for each detector. 
The photon count rate (PCR) was subsequently determined by subtracting the DCR from the total count rate.
Detectors exhibiting a saturation plateau, where the PCR remains constant despite the bias conditions, are considered to have reached unity internal detection efficiency (IDE) and their PCR was normalized to this plateau. 
We normalized the PCR from detectors that did not reach a clear saturation point to their value at the point of the lowest increasing rate. 

\subsection{Acquisition parameters}

\begin{figure}
    \includegraphics[width=1\textwidth]{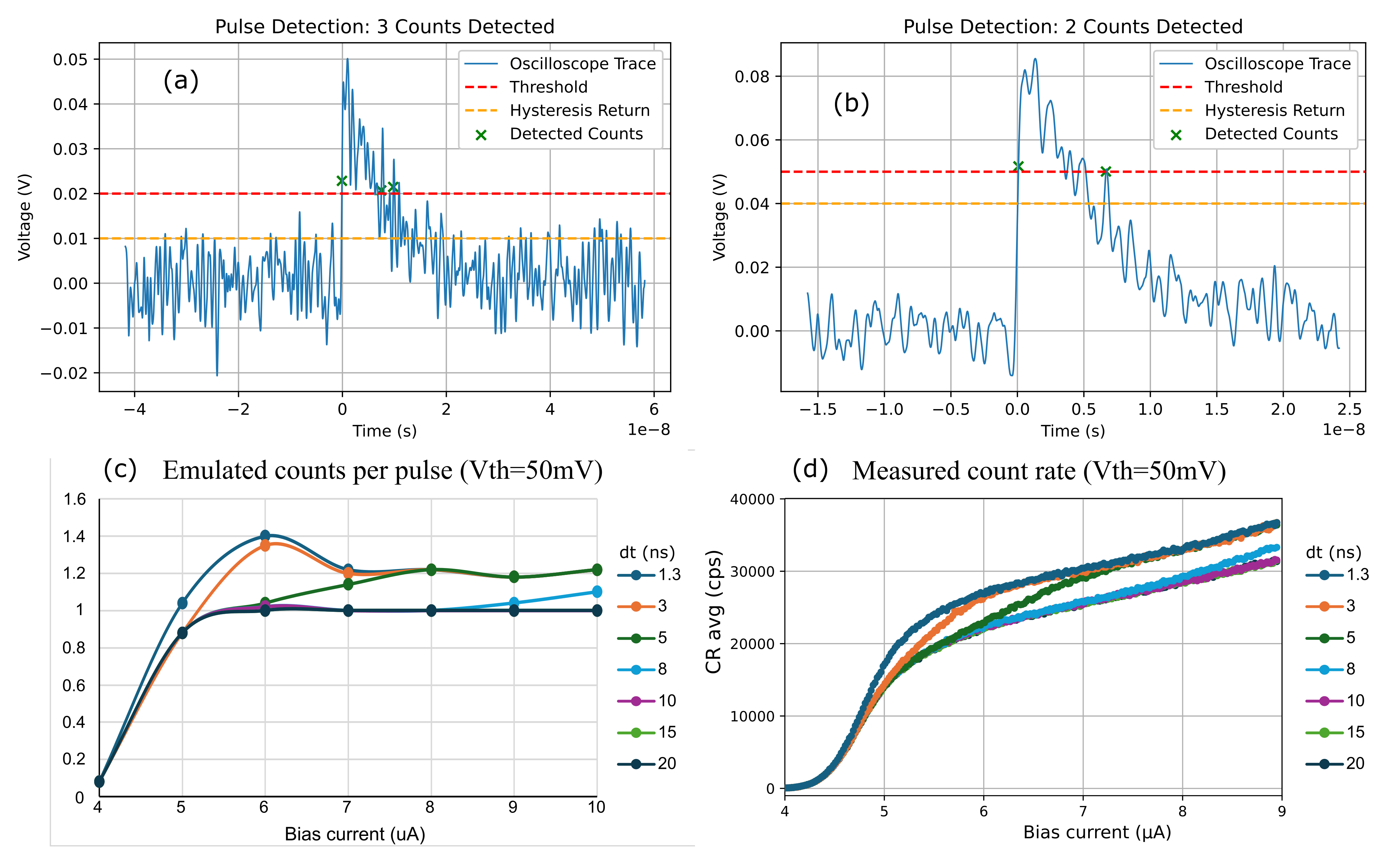}%
    \caption{Effect of acquisition parameters using an SNSPD with $W=\qty{115}{nm}$ and $t=\qty{8}{nm}$. (a-b) Emulation of counting acquisition using an oscilloscope trace, when using (a) low and (b) higher bias current and threshold value. (c) Over-counting emulation using many oscilloscope traces with varying time tagger dead-times. (d) Measured count rate using varying time tagger dead-times. }    
    \label{fig:acq_vth_dt}
\end{figure}

A time tagger (Swabian Instruments Time Tagger X) was employed to record the count rates of SNSPDs in this study. 
Over a defined timebin, the instrument assigns a precise timestamp to the averaged input pulses during this time-window (e.g., dark or total count of SNSPDs) when the they cross a user-defined discriminator threshold ($V_{th}$), enabling the reconstruction of count statistics. 
Either rising or falling threshold crossing-event can be registered. 
In this study, the selected edge was chosen according to the bias polarity.
After a registration event, two factors limit when the time tagger will be armed again for the next event.

\begin{itemize}
    \item Hysteresis: Taking registration at rising edge as the example, after an event, the time tagger is re-armed only when the signal voltage ($V_s$) has fallen sufficiently below the level defined by the discriminator threshold ($V_{th}$) and hysteresis ($V_{hys}$). For instance, $|V_s|<|V_{th}-V_{hys}|$.
    \item Acquisition dead-time $t_{d,acq}$: Following a registered event, the time tagger is insensitive for a fixed duration, before it can register another event. In our setup, the minimum $t_{d,acq}$ is \qty{1.3}{ns}.
\end{itemize}

When measuring count-rates close to the detection threshold, it is of high importance to choose the dead-time correctly.
At this bias condition, detection peak amplitudes are only slightly larger than the noise voltage, causing multiple registered events for a single physical photon detection event. 

To study the significance of the dead-time, we chose an SNSPD and measured many detection signal peaks with an oscilloscope at multiple values of $I_b$ (Fig.~\ref{fig:acq_vth_dt}(a) and (b)). 
We then emulated the behavior captured from the oscilloscope with a script. 
In parallel, for the same device, we measured the count-rate as a function of $I_b$ with the time tagger multiple times, each time setting a different $t_{d,acq}$. 
Fig.~\ref{fig:acq_vth_dt}(c) shows the emulated average counts per pulse and Fig.~\ref{fig:acq_vth_dt}(d) shows the measured count-rates.

From these observations, it is clear that the best way to ensure a single count per pulse is by setting a high $V_{th}$ and a long $t_{d,acq}$. 
However, setting a high $V_{th}$ comes at the cost of not being able to measure detection pulses at low values of $I_b$. 
A long $t_{d,acq}$ comes at the risk of not measuring subsequent legitimate detection pulses that happen during $t_{d,acq}$.

\section{Uncertainty calculations}
Errors in the $\Delta I_{det}$ and plateau width calculations were obtained using the Monte Carlo method. The assumption was made that the underlying curve of the PCR as a function of bias current is smooth and small PCR variations between data points are caused by the stochastic nature of photon absorption in the nanowire. We fit a spline to the PCR data and modeled the PCR variation as bias-current dependent variance. 
The Monte Carlo method was subsequently performed by
\begin{enumerate}
    \item randomly generating noise data based on the modeled bias current dependent variance,
    \item adding the noise data to the fitted spline,
    \item refitting splines to the new, noisy datasets,
    \item extracting the count rate at the saturation plateau for each generated dataset from the fitted spline,
    \item and finally extracting the crossing values at 10\% and 90\% of the IDE plateau from the fitted curve.
\end{enumerate}
The errors extracted from this method account for bias current-dependent variance in count rate and uncertainty in the plateau value extraction.
A similar method was used to extract the bias current value where the DCR exceeds \qty{10}{cps}.

\section{Simulations}

Figure \ref{fig:simulations}(a) and (b) show $I_c$ simulations of straight wires and the outgoing bends toward the bonding pads respectively. 
This is shown for three different widths: \qty{12}{\xi}, \qty{16}{\xi} and \qty{20}{\xi}. The outgoing bends show strong dependence of $I_c$ depending on the direction of the magnetic field (or, equivalently, the bias polarity). 

\begin{figure}
    \includegraphics[width=0.9\textwidth]{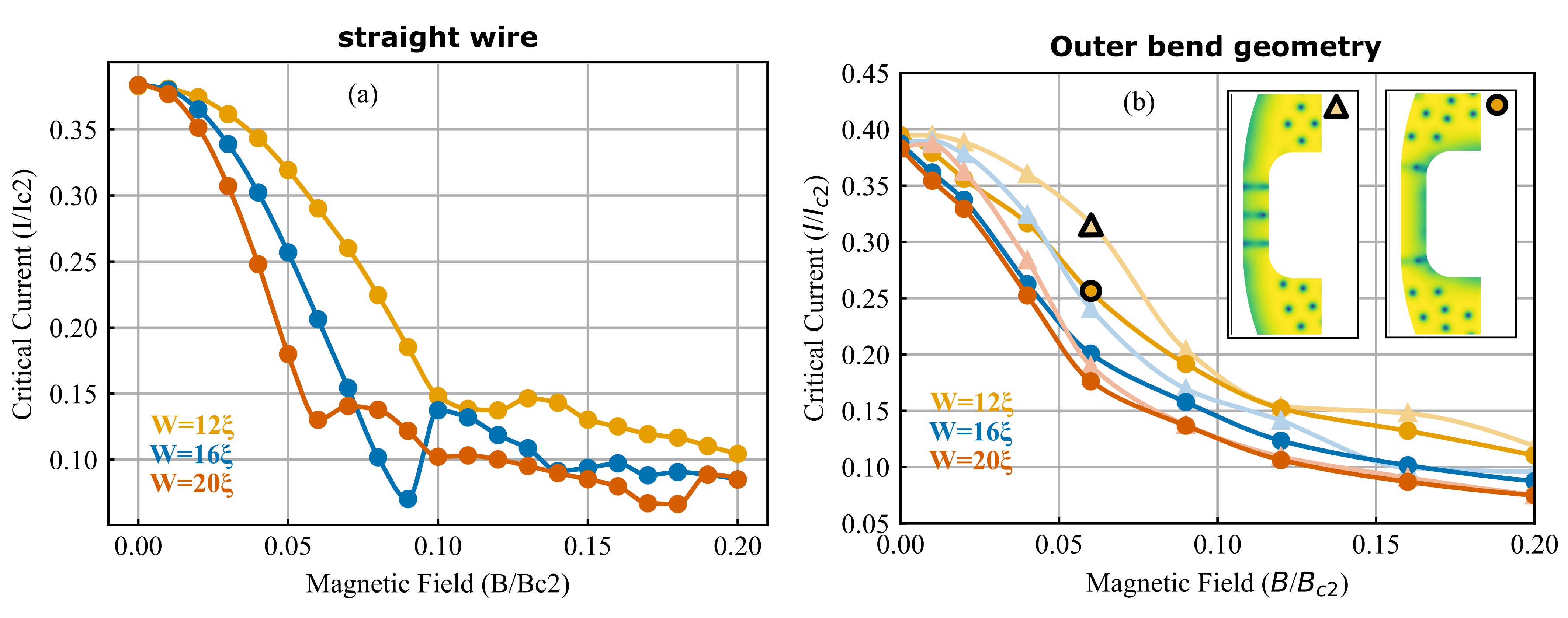}%
    \caption{Straight wire and outer bend geometry simulations.}
    
    \label{fig:simulations}
\end{figure}

Figure \ref{fig:jb_W_sim} shows the simulated bias current density over the cross-section of the nanowire, for different nanowire widths. 
In the absence of a magnetic field, the bias current is distributed uniformly over the nanowire. 
When a magnetic field is applied, perpendicular to the film orientation, the bias current varies near linearly over the width. 
The total difference in bias current density between the two nanowire edges is directly related to the width of the nanowire.

\begin{figure}
    \includegraphics[width=0.9\textwidth]{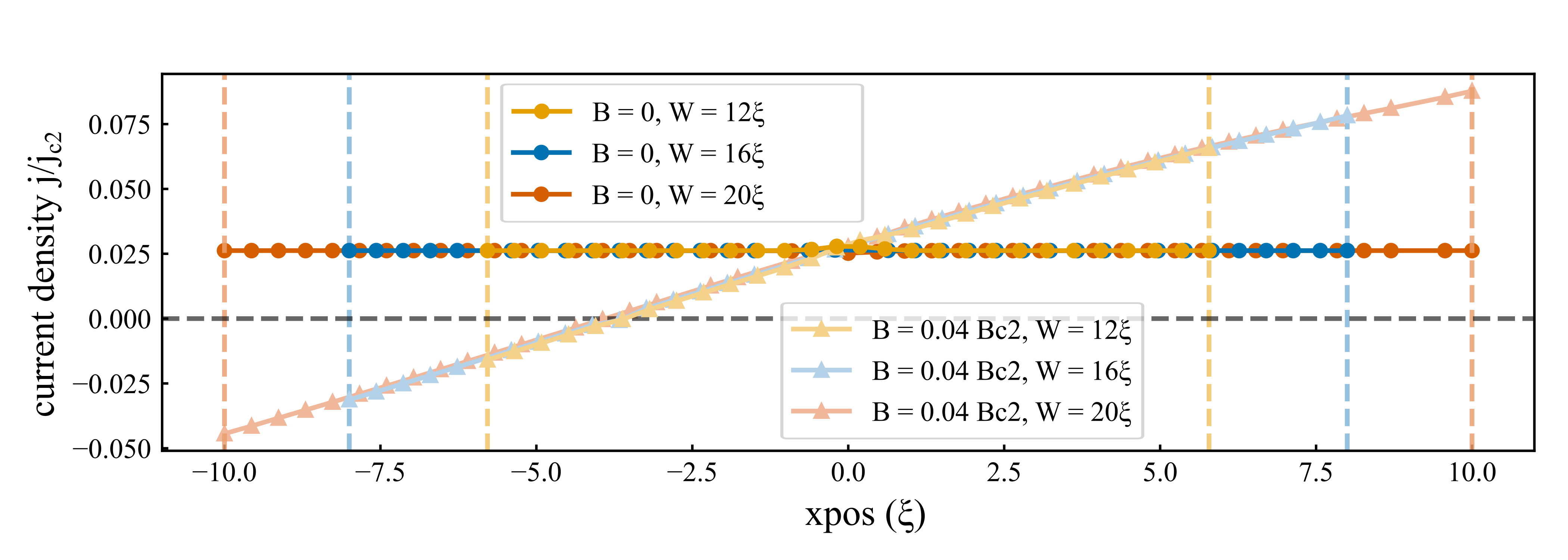}%
    \caption{Distribution of bias current over the nanowire for different widths.}
    
    \label{fig:jb_W_sim}
\end{figure}

\section{PCR data for all devices and wavelengths}

Figure \ref{fig:FOM_6nm_8nm}(a) and (b) show the saturation plateau width for the devices with layer thickness of \qty{6}{nm} and \qty{8}{nm} respectively. 

\begin{figure}
    \includegraphics[width=0.9\textwidth]{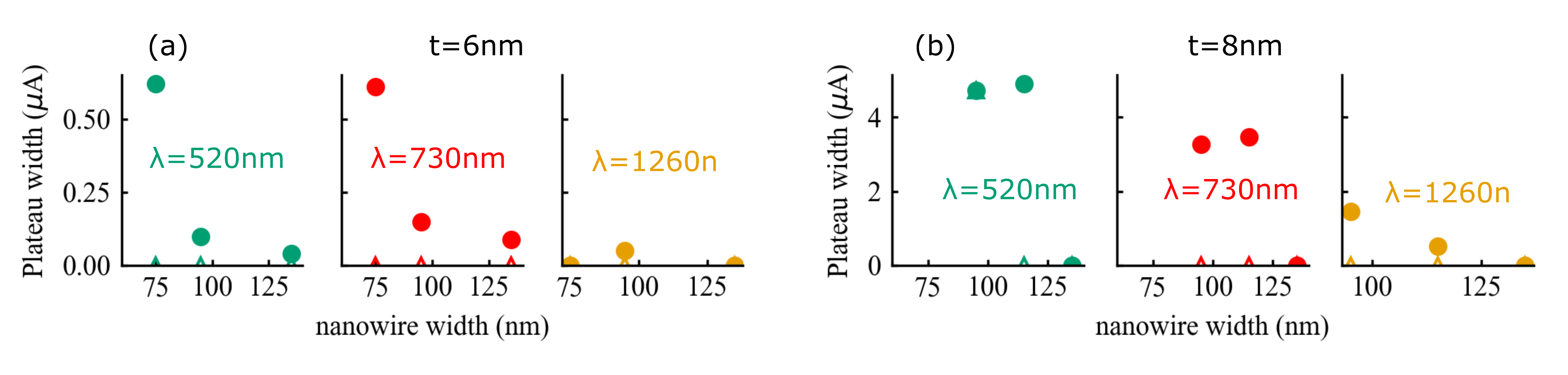}%
    \caption{(a) Plateau width for the sample with \qty{6}{nm} NbTiN thickness. (b) Plateau width for the sample with \qty{8}{nm} NbTiN thickness.}
    \label{fig:FOM_6nm_8nm}
\end{figure}

Figures \ref{fig:10nm_all_IV}, \ref{fig:8nm_all_IV} and \ref{fig:6nm_all_IV} show IV measurements of the samples with layer thickness \qty{10}{nm}, \qty{8}{nm} and \qty{6}{nm} respectively. 
Note that the voltage is measured at the ADC input (see Fig.~\ref{fig:bias_tee}, and therefore the IV relation does not depend on the series resistance. 

\begin{figure}
    \includegraphics[width=0.9\textwidth]{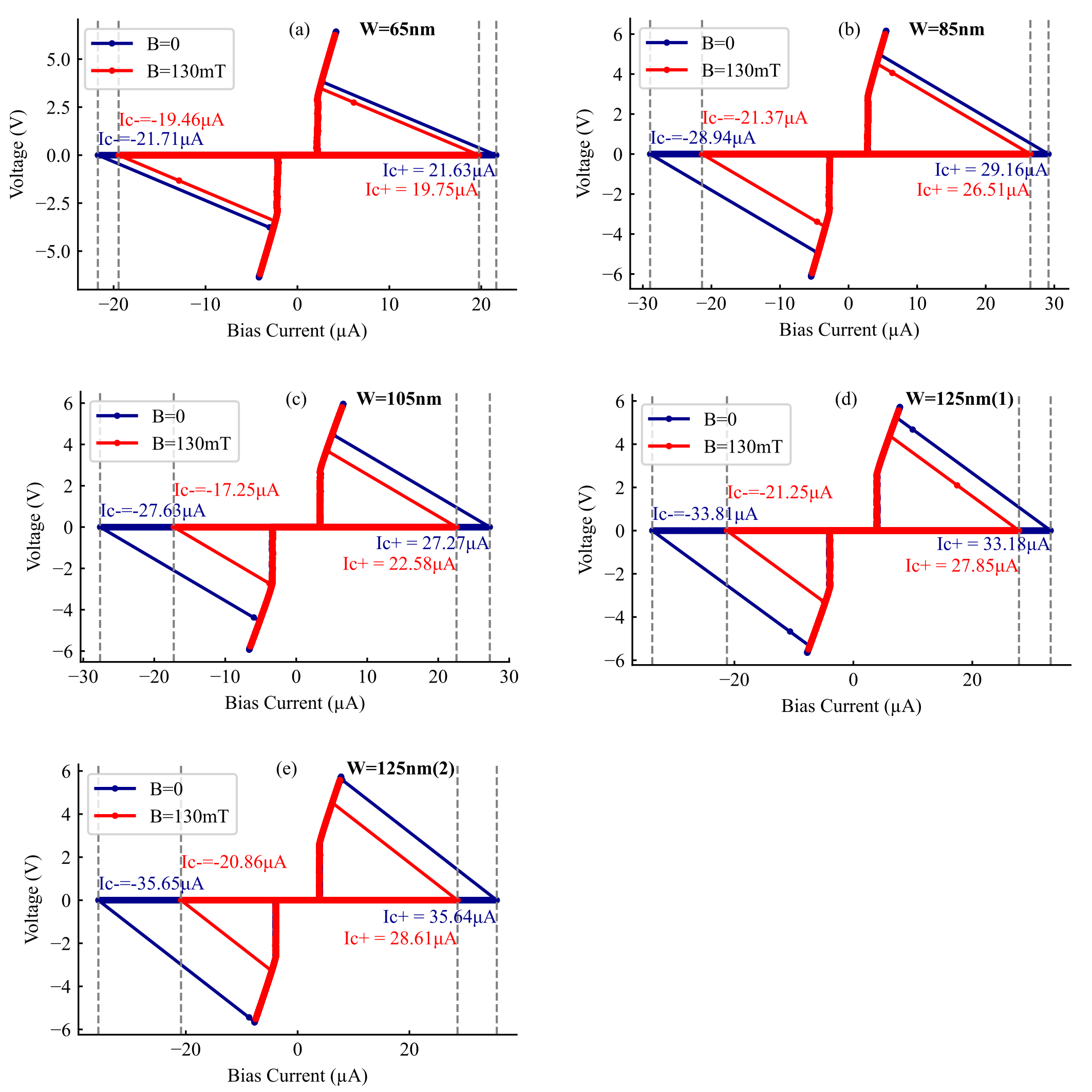}%
    \caption{IV measurements of the devices with 10~nm NbTiN thickness.}
    
    \label{fig:10nm_all_IV}
\end{figure}

\begin{figure}
    \includegraphics[width=0.9\textwidth]{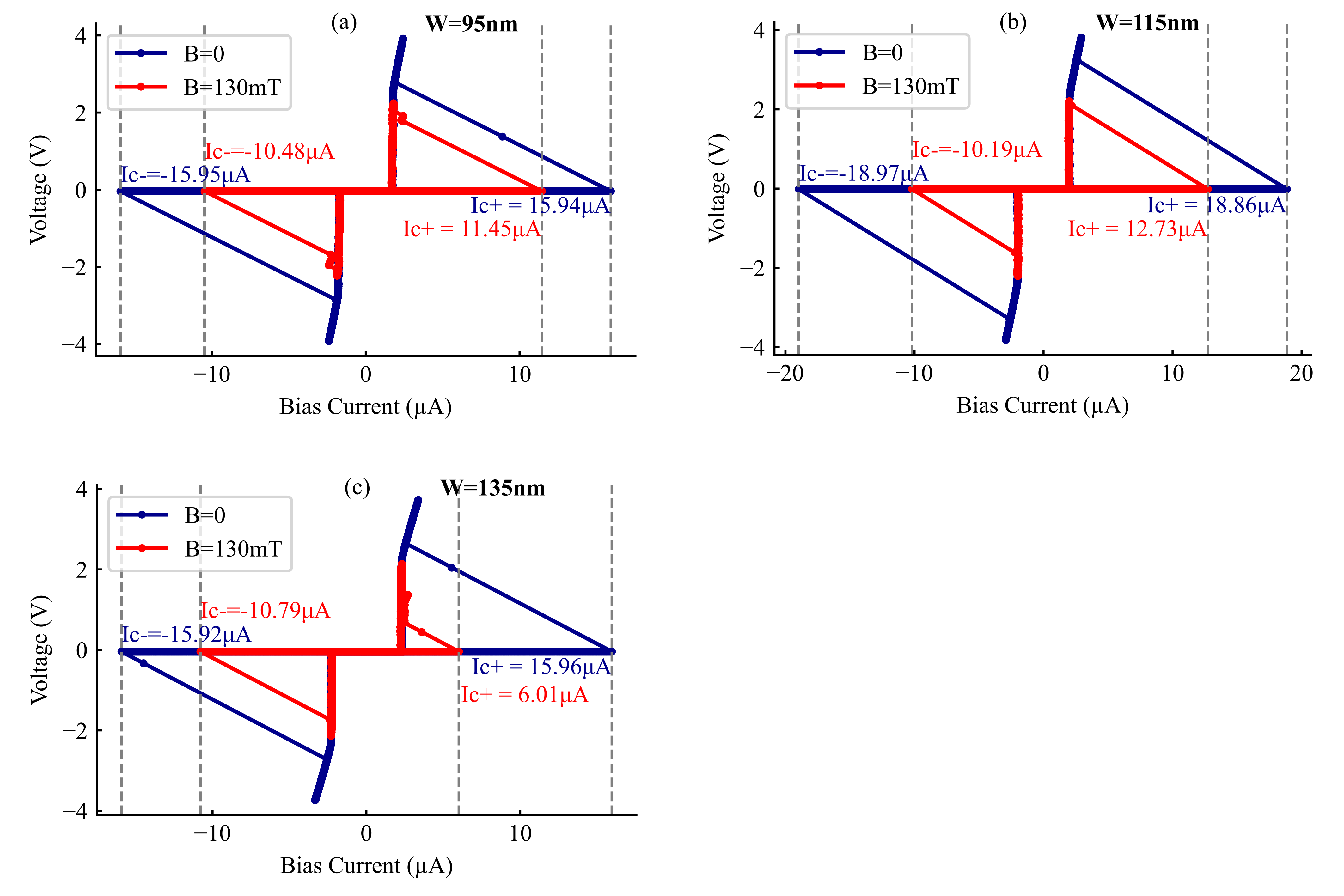}%
    \caption{IV measurements of the devices with 8~nm NbTiN thickness.}
    
    \label{fig:8nm_all_IV}
\end{figure}

\begin{figure}
    \includegraphics[width=0.9\textwidth]{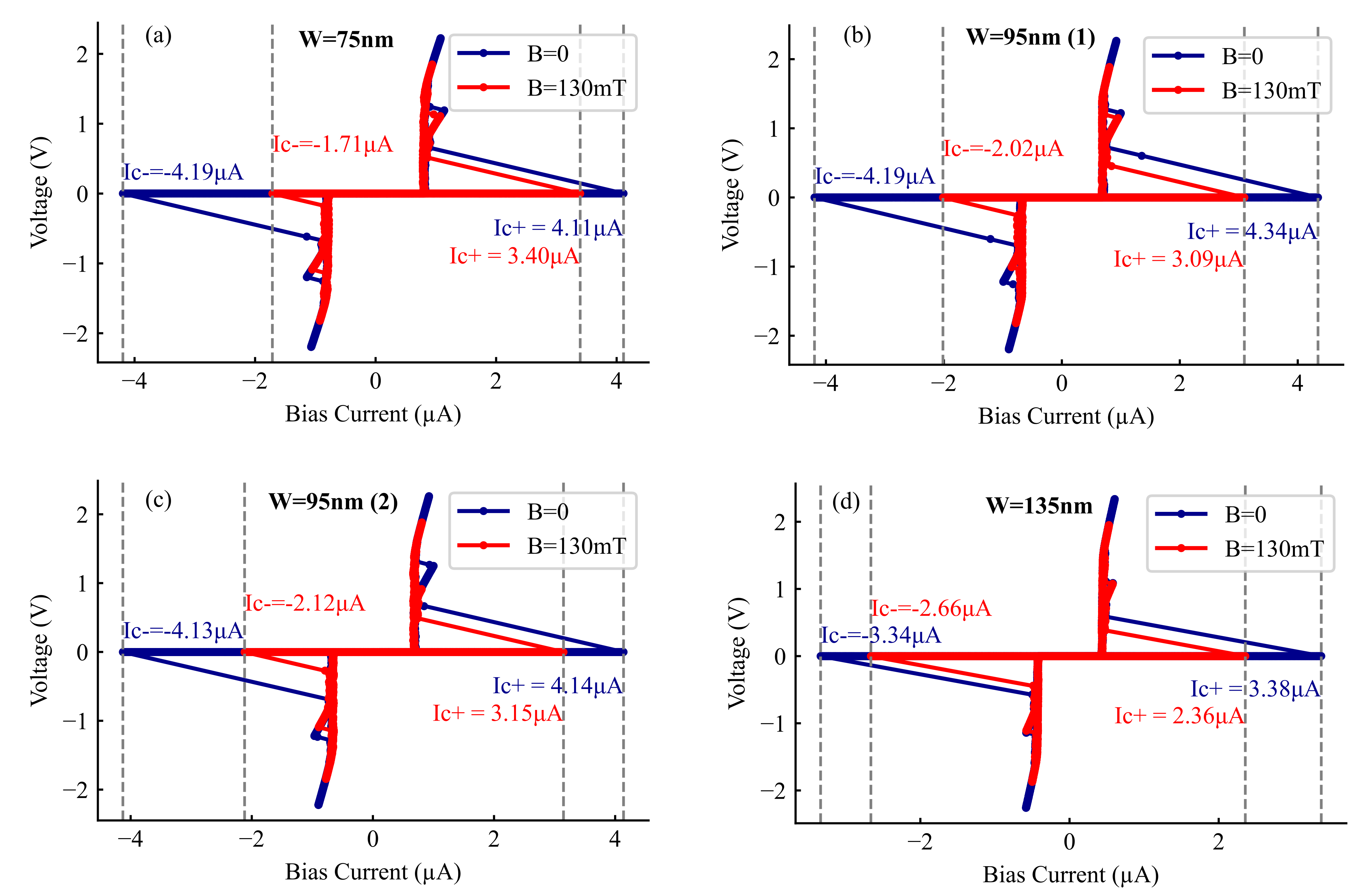}%
    \caption{IV measurements of the devices with 6~nm NbTiN thickness.}
    
    \label{fig:6nm_all_IV}
\end{figure}

Figures \ref{fig:10nm_all_pcr}, \ref{fig:8nm_all_pcr} and \ref{fig:6nm_all_pcr} show DCR and normalized PCR measurements of the samples with layer thickness \qty{10}{nm}, \qty{8}{nm} and \qty{6}{nm} respectively. 
Insets show $\Delta I_{det}$. 
Note that only datasets are shown where the magnitude of the saturation plateau can be derived.

\begin{figure}
    \includegraphics[width=0.9\textwidth]{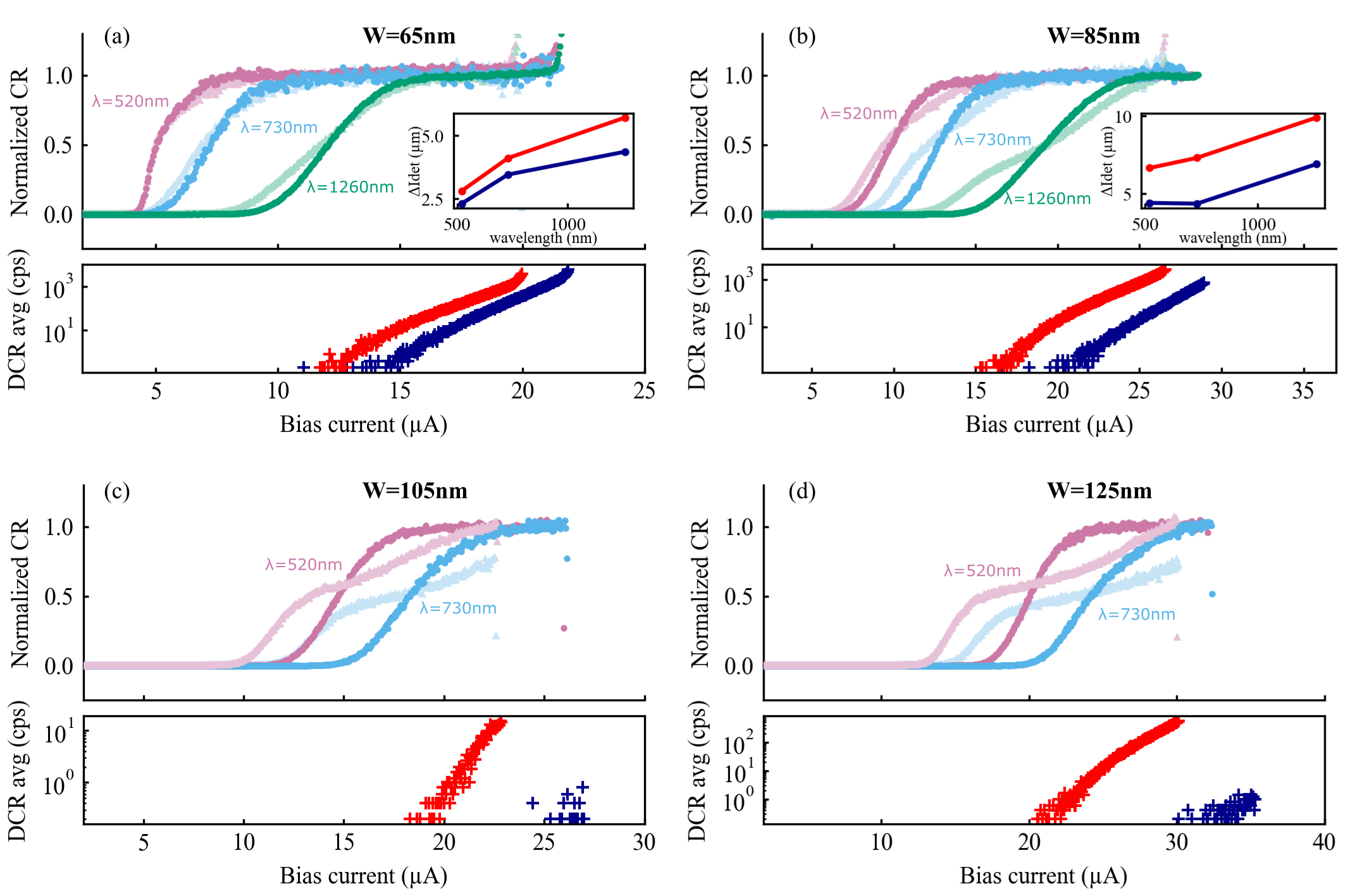}%
    \caption{PCR and DCR data of the devices with 10~nm NbTiN thickness.}
    
    \label{fig:10nm_all_pcr}
\end{figure}

\begin{figure}
    \includegraphics[width=0.9\textwidth]{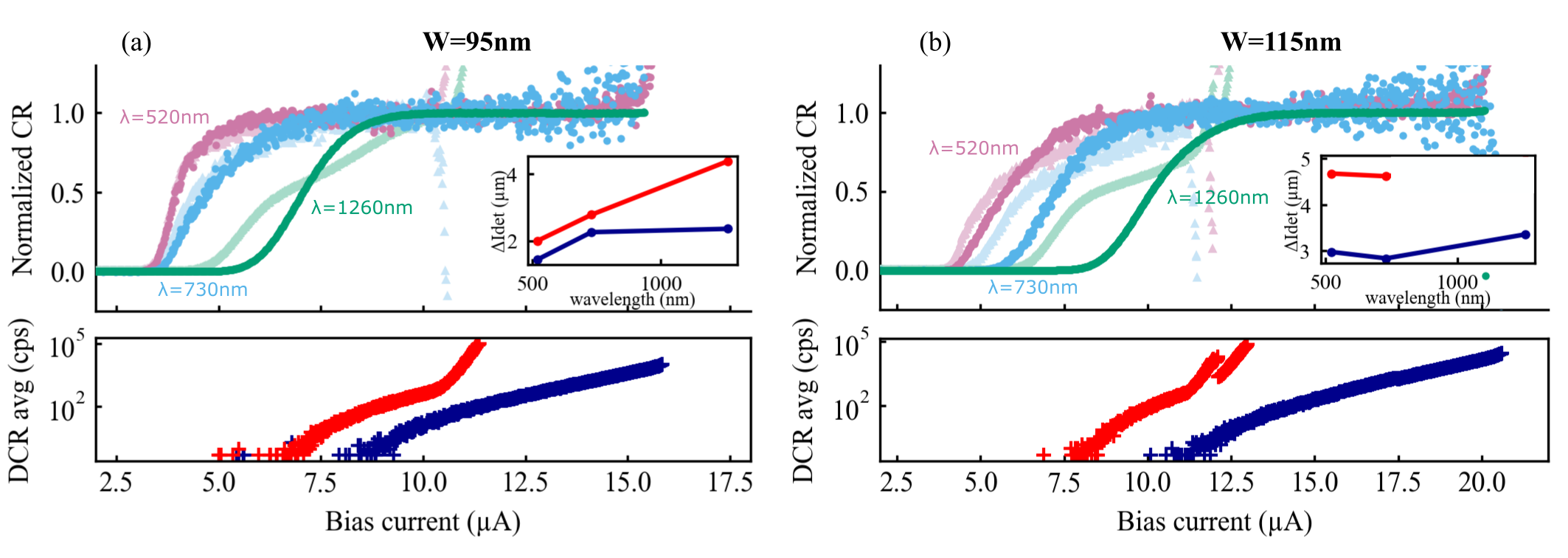}%
    \caption{PCR and DCR data of the devices with 8~nm NbTiN thickness.}
    
    \label{fig:8nm_all_pcr}
\end{figure}

\begin{figure}
    \includegraphics[width=0.9\textwidth]{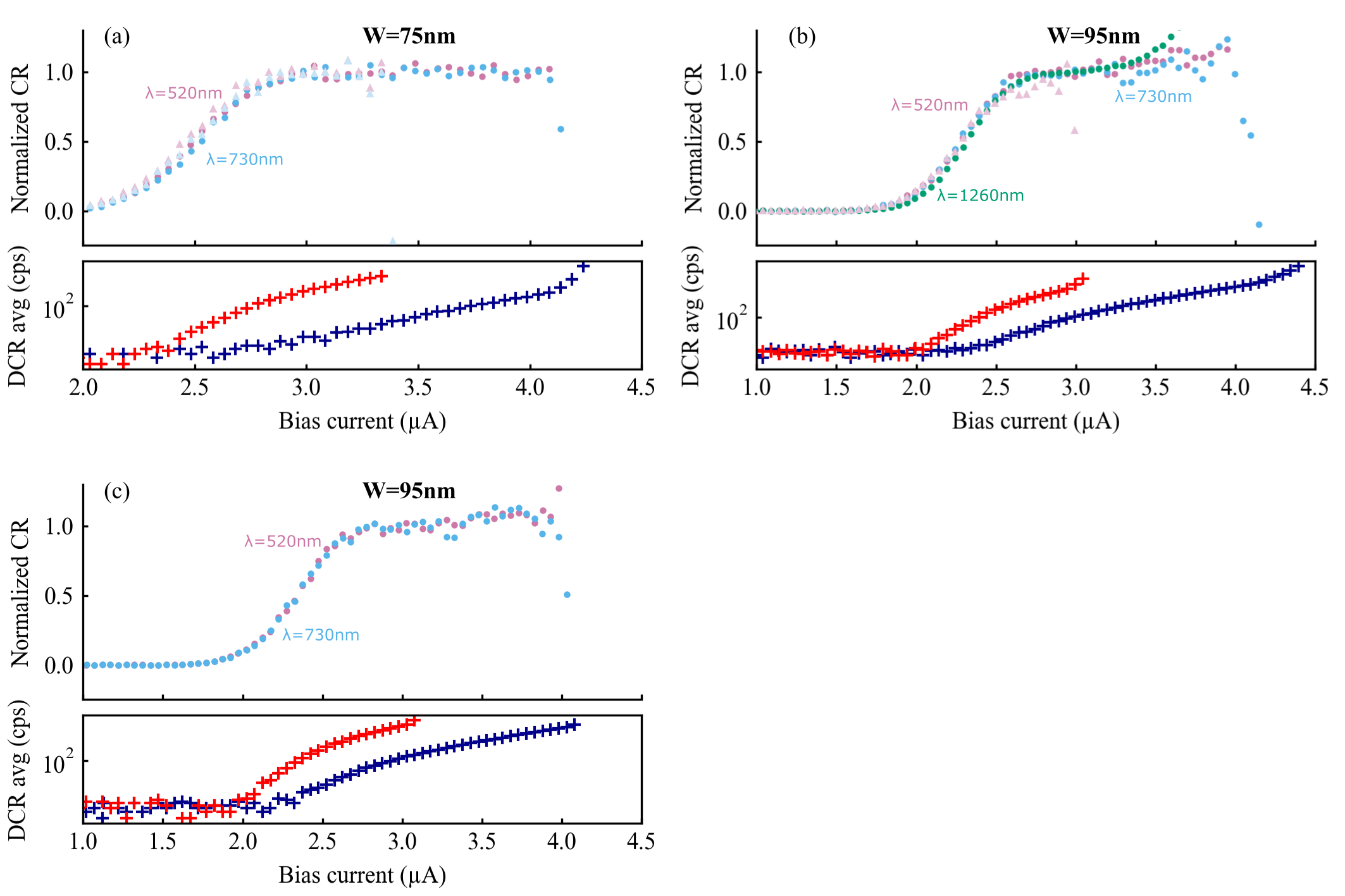}%
    \caption{PCR and DCR data of the devices with 6~nm NbTiN thickness.}
    
    \label{fig:6nm_all_pcr}
\end{figure}

\end{suppinfo}

\providecommand{\latin}[1]{#1}
\makeatletter
\providecommand{\doi}
  {\begingroup\let\do\@makeother\dospecials
  \catcode`\{=1 \catcode`\}=2 \doi@aux}
\providecommand{\doi@aux}[1]{\endgroup\texttt{#1}}
\makeatother
\providecommand*\mcitethebibliography{\thebibliography}
\csname @ifundefined\endcsname{endmcitethebibliography}  {\let\endmcitethebibliography\endthebibliography}{}


\begin{mcitethebibliography}{54}
\providecommand*\natexlab[1]{#1}
\providecommand*\mciteSetBstSublistMode[1]{}
\providecommand*\mciteSetBstMaxWidthForm[2]{}
\providecommand*\mciteBstWouldAddEndPuncttrue
  {\def\EndOfBibitem{\unskip.}}
\providecommand*\mciteBstWouldAddEndPunctfalse
  {\let\EndOfBibitem\relax}
\providecommand*\mciteSetBstMidEndSepPunct[3]{}
\providecommand*\mciteSetBstSublistLabelBeginEnd[3]{}
\providecommand*\EndOfBibitem{}
\mciteSetBstSublistMode{f}
\mciteSetBstMaxWidthForm{subitem}{(\alph{mcitesubitemcount})}
\mciteSetBstSublistLabelBeginEnd
  {\mcitemaxwidthsubitemform\space}
  {\relax}
  {\relax}

\bibitem[Awschalom \latin{et~al.}(2018)Awschalom, Hanson, Wrachtrup, and Zhou]{awschalom2018quantum}
Awschalom,~D.~D.; Hanson,~R.; Wrachtrup,~J.; Zhou,~B.~B. Quantum technologies with optically interfaced solid-state spins. \emph{Nature Photonics} \textbf{2018}, \emph{12}, 516--527\relax
\mciteBstWouldAddEndPuncttrue
\mciteSetBstMidEndSepPunct{\mcitedefaultmidpunct}
{\mcitedefaultendpunct}{\mcitedefaultseppunct}\relax
\EndOfBibitem
\bibitem[Ebadi \latin{et~al.}(2021)Ebadi, Wang, Levine, Keesling, Semeghini, Omran, Bluvstein, Samajdar, Pichler, Ho, \latin{et~al.} others]{ebadi2021quantum}
Ebadi,~S.; Wang,~T.~T.; Levine,~H.; Keesling,~A.; Semeghini,~G.; Omran,~A.; Bluvstein,~D.; Samajdar,~R.; Pichler,~H.; Ho,~W.~W.; others Quantum phases of matter on a 256-atom programmable quantum simulator. \emph{Nature} \textbf{2021}, \emph{595}, 227--232\relax
\mciteBstWouldAddEndPuncttrue
\mciteSetBstMidEndSepPunct{\mcitedefaultmidpunct}
{\mcitedefaultendpunct}{\mcitedefaultseppunct}\relax
\EndOfBibitem
\bibitem[Bruzewicz \latin{et~al.}(2019)Bruzewicz, Chiaverini, McConnell, and Sage]{bruzewicz2019trapped}
Bruzewicz,~C.~D.; Chiaverini,~J.; McConnell,~R.; Sage,~J.~M. Trapped-ion quantum computing: Progress and challenges. \emph{Applied physics reviews} \textbf{2019}, \emph{6}\relax
\mciteBstWouldAddEndPuncttrue
\mciteSetBstMidEndSepPunct{\mcitedefaultmidpunct}
{\mcitedefaultendpunct}{\mcitedefaultseppunct}\relax
\EndOfBibitem
\bibitem[Kjaergaard \latin{et~al.}(2020)Kjaergaard, Schwartz, Braum{\"u}ller, Krantz, Wang, Gustavsson, and Oliver]{kjaergaard2020superconducting}
Kjaergaard,~M.; Schwartz,~M.~E.; Braum{\"u}ller,~J.; Krantz,~P.; Wang,~J. I.-J.; Gustavsson,~S.; Oliver,~W.~D. Superconducting qubits: Current state of play. \emph{Annual Review of Condensed Matter Physics} \textbf{2020}, \emph{11}, 369--395\relax
\mciteBstWouldAddEndPuncttrue
\mciteSetBstMidEndSepPunct{\mcitedefaultmidpunct}
{\mcitedefaultendpunct}{\mcitedefaultseppunct}\relax
\EndOfBibitem
\bibitem[Koppell \latin{et~al.}(2025)Koppell, Bittencourt, Paul, Huang, Baryakhtar, and Berggren]{koppell2025dark}
Koppell,~S.; Bittencourt,~O.~D.; Paul,~D.~J.; Huang,~J.; Baryakhtar,~M.; Berggren,~K.~K. Dark Matter Haloscope with a Disordered Dielectric Absorber. \emph{arXiv preprint arXiv:2506.00115} \textbf{2025}, \relax
\mciteBstWouldAddEndPunctfalse
\mciteSetBstMidEndSepPunct{\mcitedefaultmidpunct}
{}{\mcitedefaultseppunct}\relax
\EndOfBibitem
\bibitem[Armstrong \latin{et~al.}(2015)Armstrong, Choi, Kaczanowicz, Lukhanin, Meziani, and Sawatzky]{armstrong2015threshold}
Armstrong,~W.~R.; Choi,~S.; Kaczanowicz,~E.; Lukhanin,~A.; Meziani,~Z.-E.; Sawatzky,~B. A threshold gas Cherenkov detector for the Spin Asymmetries of the Nucleon Experiment. \emph{Nuclear Instruments and Methods in Physics Research Section A: Accelerators, Spectrometers, Detectors and Associated Equipment} \textbf{2015}, \emph{804}, 118--126\relax
\mciteBstWouldAddEndPuncttrue
\mciteSetBstMidEndSepPunct{\mcitedefaultmidpunct}
{\mcitedefaultendpunct}{\mcitedefaultseppunct}\relax
\EndOfBibitem
\bibitem[Polakovic \latin{et~al.}(2020)Polakovic, Armstrong, Yefremenko, Pearson, Hafidi, Karapetrov, Meziani, and Novosad]{polakovic_superconducting_2020}
Polakovic,~T.; Armstrong,~W.~R.; Yefremenko,~V.; Pearson,~J.~E.; Hafidi,~K.; Karapetrov,~G.; Meziani,~Z.~E.; Novosad,~V. Superconducting nanowires as high-rate photon detectors in strong magnetic fields. \emph{Nuclear Instruments and Methods in Physics Research Section A: Accelerators, Spectrometers, Detectors and Associated Equipment} \textbf{2020}, \emph{959}, 163543\relax
\mciteBstWouldAddEndPuncttrue
\mciteSetBstMidEndSepPunct{\mcitedefaultmidpunct}
{\mcitedefaultendpunct}{\mcitedefaultseppunct}\relax
\EndOfBibitem
\bibitem[goo(2025)]{google2025quantum}
Quantum error correction below the surface code threshold. \emph{Nature} \textbf{2025}, \emph{638}, 920--926\relax
\mciteBstWouldAddEndPuncttrue
\mciteSetBstMidEndSepPunct{\mcitedefaultmidpunct}
{\mcitedefaultendpunct}{\mcitedefaultseppunct}\relax
\EndOfBibitem
\bibitem[Venza and Colangelo(2025)Venza, and Colangelo]{venza2025research}
Venza,~F.~P.; Colangelo,~M. Research trends in single-photon detectors based on superconducting wires. \emph{APL Photonics} \textbf{2025}, \emph{10}\relax
\mciteBstWouldAddEndPuncttrue
\mciteSetBstMidEndSepPunct{\mcitedefaultmidpunct}
{\mcitedefaultendpunct}{\mcitedefaultseppunct}\relax
\EndOfBibitem
\bibitem[Reddy \latin{et~al.}(2020)Reddy, Nerem, Nam, Mirin, and Verma]{reddy2020superconducting}
Reddy,~D.~V.; Nerem,~R.~R.; Nam,~S.~W.; Mirin,~R.~P.; Verma,~V.~B. Superconducting nanowire single-photon detectors with 98\% system detection efficiency at 1550 nm. \emph{Optica} \textbf{2020}, \emph{7}, 1649--1653\relax
\mciteBstWouldAddEndPuncttrue
\mciteSetBstMidEndSepPunct{\mcitedefaultmidpunct}
{\mcitedefaultendpunct}{\mcitedefaultseppunct}\relax
\EndOfBibitem
\bibitem[Verma \latin{et~al.}(2021)Verma, Korzh, Walter, Lita, Briggs, Colangelo, Zhai, Wollman, Beyer, Allmaras, \latin{et~al.} others]{verma2021single}
Verma,~V.; Korzh,~B.; Walter,~A.~B.; Lita,~A.~E.; Briggs,~R.~M.; Colangelo,~M.; Zhai,~Y.; Wollman,~E.~E.; Beyer,~A.~D.; Allmaras,~J.~P.; others Single-photon detection in the mid-infrared up to 10 $\mu$m wavelength using tungsten silicide superconducting nanowire detectors. \emph{APL photonics} \textbf{2021}, \emph{6}\relax
\mciteBstWouldAddEndPuncttrue
\mciteSetBstMidEndSepPunct{\mcitedefaultmidpunct}
{\mcitedefaultendpunct}{\mcitedefaultseppunct}\relax
\EndOfBibitem
\bibitem[Korzh \latin{et~al.}(2020)Korzh, Zhao, Allmaras, Frasca, Autry, Bersin, Beyer, Briggs, Bumble, Colangelo, \latin{et~al.} others]{korzh2020demonstration}
Korzh,~B.; Zhao,~Q.-Y.; Allmaras,~J.~P.; Frasca,~S.; Autry,~T.~M.; Bersin,~E.~A.; Beyer,~A.~D.; Briggs,~R.~M.; Bumble,~B.; Colangelo,~M.; others Demonstration of sub-3 ps temporal resolution with a superconducting nanowire single-photon detector. \emph{Nature Photonics} \textbf{2020}, \emph{14}, 250--255\relax
\mciteBstWouldAddEndPuncttrue
\mciteSetBstMidEndSepPunct{\mcitedefaultmidpunct}
{\mcitedefaultendpunct}{\mcitedefaultseppunct}\relax
\EndOfBibitem
\bibitem[Ferrari \latin{et~al.}(2018)Ferrari, Schuck, and Pernice]{ferrari_waveguide-integrated_2018}
Ferrari,~S.; Schuck,~C.; Pernice,~W. Waveguide-integrated superconducting nanowire single-photon detectors. \emph{Nanophotonics} \textbf{2018}, \emph{7}, 1725--1758\relax
\mciteBstWouldAddEndPuncttrue
\mciteSetBstMidEndSepPunct{\mcitedefaultmidpunct}
{\mcitedefaultendpunct}{\mcitedefaultseppunct}\relax
\EndOfBibitem
\bibitem[Degen \latin{et~al.}(2017)Degen, Reinhard, and Cappellaro]{degen2017quantum}
Degen,~C.~L.; Reinhard,~F.; Cappellaro,~P. Quantum sensing. \emph{Reviews of modern physics} \textbf{2017}, \emph{89}, 035002\relax
\mciteBstWouldAddEndPuncttrue
\mciteSetBstMidEndSepPunct{\mcitedefaultmidpunct}
{\mcitedefaultendpunct}{\mcitedefaultseppunct}\relax
\EndOfBibitem
\bibitem[Barry \latin{et~al.}(2020)Barry, Schloss, Bauch, Turner, Hart, Pham, and Walsworth]{barry2020sensitivity}
Barry,~J.~F.; Schloss,~J.~M.; Bauch,~E.; Turner,~M.~J.; Hart,~C.~A.; Pham,~L.~M.; Walsworth,~R.~L. Sensitivity optimization for NV-diamond magnetometry. \emph{Reviews of modern physics} \textbf{2020}, \emph{92}, 015004\relax
\mciteBstWouldAddEndPuncttrue
\mciteSetBstMidEndSepPunct{\mcitedefaultmidpunct}
{\mcitedefaultendpunct}{\mcitedefaultseppunct}\relax
\EndOfBibitem
\bibitem[Hortensius \latin{et~al.}(2012)Hortensius, Driessen, Klapwijk, Berggren, and Clem]{hortensius2012critical}
Hortensius,~H.; Driessen,~E.; Klapwijk,~T.; Berggren,~K.; Clem,~J. Critical-current reduction in thin superconducting wires due to current crowding. \emph{Applied Physics Letters} \textbf{2012}, \emph{100}\relax
\mciteBstWouldAddEndPuncttrue
\mciteSetBstMidEndSepPunct{\mcitedefaultmidpunct}
{\mcitedefaultendpunct}{\mcitedefaultseppunct}\relax
\EndOfBibitem
\bibitem[Henrich \latin{et~al.}(2012)Henrich, Reichensperger, Hofherr, Meckbach, Il'in, Siegel, Semenov, Zotova, and Vodolazov]{henrich2012geometry}
Henrich,~D.; Reichensperger,~P.; Hofherr,~M.; Meckbach,~J.; Il'in,~K.; Siegel,~M.; Semenov,~A.; Zotova,~A.; Vodolazov,~D.~Y. Geometry-induced reduction of the critical current in superconducting nanowires. \emph{Physical Review B—Condensed Matter and Materials Physics} \textbf{2012}, \emph{86}, 144504\relax
\mciteBstWouldAddEndPuncttrue
\mciteSetBstMidEndSepPunct{\mcitedefaultmidpunct}
{\mcitedefaultendpunct}{\mcitedefaultseppunct}\relax
\EndOfBibitem
\bibitem[Semenov \latin{et~al.}(2015)Semenov, Charaev, Lusche, Ilin, Siegel, Hübers, Bralović, Dopf, and Vodolazov]{semenov_asymmetry_2015}
Semenov,~A.; Charaev,~I.; Lusche,~R.; Ilin,~K.; Siegel,~M.; Hübers,~H.-W.; Bralović,~N.; Dopf,~K.; Vodolazov,~D.~Y. Asymmetry in the effect of magnetic field on photon detection and dark counts in bended nanostrips. \emph{Physical Review B} \textbf{2015}, \emph{92}, 174518\relax
\mciteBstWouldAddEndPuncttrue
\mciteSetBstMidEndSepPunct{\mcitedefaultmidpunct}
{\mcitedefaultendpunct}{\mcitedefaultseppunct}\relax
\EndOfBibitem
\bibitem[Kerman \latin{et~al.}(2007)Kerman, Dauler, Yang, Rosfjord, Anant, Berggren, Gol’tsman, and Voronov]{kerman2007constriction}
Kerman,~A.~J.; Dauler,~E.~A.; Yang,~J.~K.; Rosfjord,~K.~M.; Anant,~V.; Berggren,~K.~K.; Gol’tsman,~G.~N.; Voronov,~B.~M. Constriction-limited detection efficiency of superconducting nanowire single-photon detectors. \emph{Applied Physics Letters} \textbf{2007}, \emph{90}\relax
\mciteBstWouldAddEndPuncttrue
\mciteSetBstMidEndSepPunct{\mcitedefaultmidpunct}
{\mcitedefaultendpunct}{\mcitedefaultseppunct}\relax
\EndOfBibitem
\bibitem[Chen \latin{et~al.}(2023)Chen, Chen, Vedin, J{\"o}nsson, Gyger, Steinhauer, Lin, Chang, Lidmar, Zwiller, \latin{et~al.} others]{chen2023visualizing}
Chen,~P.-J.; Chen,~G.-H.; Vedin,~R.; J{\"o}nsson,~M.; Gyger,~S.; Steinhauer,~S.; Lin,~J.-J.; Chang,~W.-H.; Lidmar,~J.; Zwiller,~V.; others Visualizing local superconductivity of nbtin nanowires to probe inhomogeneity in single-photon detectors. \emph{ACS Applied Optical Materials} \textbf{2023}, \emph{2}, 68--75\relax
\mciteBstWouldAddEndPuncttrue
\mciteSetBstMidEndSepPunct{\mcitedefaultmidpunct}
{\mcitedefaultendpunct}{\mcitedefaultseppunct}\relax
\EndOfBibitem
\bibitem[Charaev \latin{et~al.}(2017)Charaev, Silbernagel, Bachowsky, Kuzmin, Doerner, Ilin, Semenov, Roditchev, Vodolazov, and Siegel]{charaev2017enhancement}
Charaev,~I.; Silbernagel,~T.; Bachowsky,~B.; Kuzmin,~A.; Doerner,~S.; Ilin,~K.; Semenov,~A.; Roditchev,~D.; Vodolazov,~D.~Y.; Siegel,~M. Enhancement of superconductivity in NbN nanowires by negative electron-beam lithography with positive resist. \emph{Journal of Applied Physics} \textbf{2017}, \emph{122}\relax
\mciteBstWouldAddEndPuncttrue
\mciteSetBstMidEndSepPunct{\mcitedefaultmidpunct}
{\mcitedefaultendpunct}{\mcitedefaultseppunct}\relax
\EndOfBibitem
\bibitem[Liu \latin{et~al.}(2025)Liu, Zhao, Hao, Huang, Deng, Yang, Ru, Liu, Wang, Lv, \latin{et~al.} others]{liu2025revealing}
Liu,~Z.; Zhao,~Q.-Y.; Hao,~H.; Huang,~Y.-H.; Deng,~J.; Yang,~F.; Ru,~S.-Y.; Liu,~N.-T.; Wang,~S.-H.; Lv,~K.-H.; others Revealing Nanoscale Inhomogeneities in a Superconducting Nanowire through Self-Heating Hotspot Scanning and Mapping. \emph{Nano Letters} \textbf{2025}, \emph{25}, 14175--14184\relax
\mciteBstWouldAddEndPuncttrue
\mciteSetBstMidEndSepPunct{\mcitedefaultmidpunct}
{\mcitedefaultendpunct}{\mcitedefaultseppunct}\relax
\EndOfBibitem
\bibitem[Zhang \latin{et~al.}(2018)Zhang, You, Yang, Wu, Lv, Guo, Zhang, Li, Peng, Wang, \latin{et~al.} others]{zhang2018hotspot}
Zhang,~L.; You,~L.; Yang,~X.; Wu,~J.; Lv,~C.; Guo,~Q.; Zhang,~W.; Li,~H.; Peng,~W.; Wang,~Z.; others Hotspot relaxation time of NbN superconducting nanowire single-photon detectors on various substrates. \emph{Scientific reports} \textbf{2018}, \emph{8}, 1486\relax
\mciteBstWouldAddEndPuncttrue
\mciteSetBstMidEndSepPunct{\mcitedefaultmidpunct}
{\mcitedefaultendpunct}{\mcitedefaultseppunct}\relax
\EndOfBibitem
\bibitem[Klimov \latin{et~al.}(2017)Klimov, Słysz, Guziewicz, Kolkovsky, Zaytseva, and Malinowski]{klimov_characterization_2017}
Klimov,~A.; Słysz,~W.; Guziewicz,~M.; Kolkovsky,~V.; Zaytseva,~I.; Malinowski,~A. Characterization of the critical current and physical properties of superconducting epitaxial {NbTiN} sub-micron structures. \emph{Physica C: Superconductivity and its Applications} \textbf{2017}, \emph{536}, 35--38\relax
\mciteBstWouldAddEndPuncttrue
\mciteSetBstMidEndSepPunct{\mcitedefaultmidpunct}
{\mcitedefaultendpunct}{\mcitedefaultseppunct}\relax
\EndOfBibitem
\bibitem[Ejrnaes \latin{et~al.}(2019)Ejrnaes, Salvoni, Parlato, Massarotti, Caruso, Tafuri, Yang, You, Wang, Pepe, \latin{et~al.} others]{ejrnaes2019superconductor}
Ejrnaes,~M.; Salvoni,~D.; Parlato,~L.; Massarotti,~D.; Caruso,~R.; Tafuri,~F.; Yang,~X.; You,~L.; Wang,~Z.; Pepe,~G.; others Superconductor to resistive state switching by multiple fluctuation events in NbTiN nanostrips. \emph{Scientific reports} \textbf{2019}, \emph{9}, 8053\relax
\mciteBstWouldAddEndPuncttrue
\mciteSetBstMidEndSepPunct{\mcitedefaultmidpunct}
{\mcitedefaultendpunct}{\mcitedefaultseppunct}\relax
\EndOfBibitem
\bibitem[Il'in \latin{et~al.}(2005)Il'in, Siegel, Semenov, Engel, and Hübers]{ilin_critical_2005}
Il'in,~K.; Siegel,~M.; Semenov,~A.; Engel,~A.; Hübers,~H.-W. Critical current of Nb and {NbN} thin-film structures: The cross-section dependence. \emph{physica status solidi (c)} \textbf{2005}, \emph{2}, 1680--1687\relax
\mciteBstWouldAddEndPuncttrue
\mciteSetBstMidEndSepPunct{\mcitedefaultmidpunct}
{\mcitedefaultendpunct}{\mcitedefaultseppunct}\relax
\EndOfBibitem
\bibitem[Il’in \latin{et~al.}(2010)Il’in, Rall, Siegel, Engel, Schilling, Semenov, and Huebers]{ilin_influence_2010}
Il’in,~K.; Rall,~D.; Siegel,~M.; Engel,~A.; Schilling,~A.; Semenov,~A.; Huebers,~H.~W. Influence of thickness, width and temperature on critical current density of Nb thin film structures. \emph{Physica C: Superconductivity} \textbf{2010}, \emph{470}, 953--956\relax
\mciteBstWouldAddEndPuncttrue
\mciteSetBstMidEndSepPunct{\mcitedefaultmidpunct}
{\mcitedefaultendpunct}{\mcitedefaultseppunct}\relax
\EndOfBibitem
\bibitem[Bulaevskii \latin{et~al.}(2011)Bulaevskii, Graf, Batista, and Kogan]{bulaevskii2011vortex}
Bulaevskii,~L.; Graf,~M.; Batista,~C.; Kogan,~V. Vortex-induced dissipation in narrow current-biased thin-film superconducting strips. \emph{Physical Review B—Condensed Matter and Materials Physics} \textbf{2011}, \emph{83}, 144526\relax
\mciteBstWouldAddEndPuncttrue
\mciteSetBstMidEndSepPunct{\mcitedefaultmidpunct}
{\mcitedefaultendpunct}{\mcitedefaultseppunct}\relax
\EndOfBibitem
\bibitem[Vodolazov \latin{et~al.}(2015)Vodolazov, Korneeva, Semenov, Korneev, and Goltsman]{vodolazov_vortex-assisted_2015}
Vodolazov,~D.~Y.; Korneeva,~Y.~P.; Semenov,~A.~V.; Korneev,~A.~A.; Goltsman,~G.~N. Vortex-assisted mechanism of photon counting in a superconducting nanowire single-photon detector revealed by external magnetic field. \emph{Physical Review B} \textbf{2015}, \emph{92}, 104503\relax
\mciteBstWouldAddEndPuncttrue
\mciteSetBstMidEndSepPunct{\mcitedefaultmidpunct}
{\mcitedefaultendpunct}{\mcitedefaultseppunct}\relax
\EndOfBibitem
\bibitem[Korneeva \latin{et~al.}(2020)Korneeva, Manova, Florya, Mikhailov, Dobrovolskiy, Korneev, and Vodolazov]{korneeva2020different}
Korneeva,~Y.~P.; Manova,~N.; Florya,~I.; Mikhailov,~M.~Y.; Dobrovolskiy,~O.; Korneev,~A.; Vodolazov,~D.~Y. Different single-photon response of wide and narrow superconducting mo x si 1- x strips. \emph{Physical Review Applied} \textbf{2020}, \emph{13}, 024011\relax
\mciteBstWouldAddEndPuncttrue
\mciteSetBstMidEndSepPunct{\mcitedefaultmidpunct}
{\mcitedefaultendpunct}{\mcitedefaultseppunct}\relax
\EndOfBibitem
\bibitem[Clem and Berggren(2011)Clem, and Berggren]{clem_geometry-dependent_2011}
Clem,~J.~R.; Berggren,~K.~K. Geometry-dependent critical currents in superconducting nanocircuits. \emph{Physical Review B} \textbf{2011}, \emph{84}, 174510\relax
\mciteBstWouldAddEndPuncttrue
\mciteSetBstMidEndSepPunct{\mcitedefaultmidpunct}
{\mcitedefaultendpunct}{\mcitedefaultseppunct}\relax
\EndOfBibitem
\bibitem[J{\"o}nsson \latin{et~al.}(2022)J{\"o}nsson, Vedin, Gyger, Sutton, Steinhauer, Zwiller, Wallin, and Lidmar]{jonsson_current_2022}
J{\"o}nsson,~M.; Vedin,~R.; Gyger,~S.; Sutton,~J.~A.; Steinhauer,~S.; Zwiller,~V.; Wallin,~M.; Lidmar,~J. Current Crowding in Nanoscale Superconductors within the Ginzburg-Landau Model. \emph{Physical Review Applied} \textbf{2022}, \emph{17}, 064046\relax
\mciteBstWouldAddEndPuncttrue
\mciteSetBstMidEndSepPunct{\mcitedefaultmidpunct}
{\mcitedefaultendpunct}{\mcitedefaultseppunct}\relax
\EndOfBibitem
\bibitem[Henrich \latin{et~al.}(2013)Henrich, Rehm, D{\"o}rner, Hofherr, Il'In, Semenov, and Siegel]{henrich2013detection}
Henrich,~D.; Rehm,~L.; D{\"o}rner,~S.; Hofherr,~M.; Il'In,~K.; Semenov,~A.; Siegel,~M. Detection efficiency of a spiral-nanowire superconducting single-photon detector. \emph{IEEE transactions on applied superconductivity} \textbf{2013}, \emph{23}, 2200405--2200405\relax
\mciteBstWouldAddEndPuncttrue
\mciteSetBstMidEndSepPunct{\mcitedefaultmidpunct}
{\mcitedefaultendpunct}{\mcitedefaultseppunct}\relax
\EndOfBibitem
\bibitem[Charaev \latin{et~al.}(2019)Charaev, Semenov, Ilin, and Siegel]{charaev_magnetic-field_2019}
Charaev,~I.; Semenov,~A.; Ilin,~K.; Siegel,~M. Magnetic-Field Enhancement of Performance of Superconducting Nanowire Single-Photon Detector. \emph{{IEEE} Transactions on Applied Superconductivity} \textbf{2019}, \emph{29}, 1--5\relax
\mciteBstWouldAddEndPuncttrue
\mciteSetBstMidEndSepPunct{\mcitedefaultmidpunct}
{\mcitedefaultendpunct}{\mcitedefaultseppunct}\relax
\EndOfBibitem
\bibitem[Pearl(1964)]{pearl1964current}
Pearl,~J. Current distribution in superconducting films carrying quantized fluxoids. \emph{Applied Physics Letters} \textbf{1964}, \emph{5}, 65\relax
\mciteBstWouldAddEndPuncttrue
\mciteSetBstMidEndSepPunct{\mcitedefaultmidpunct}
{\mcitedefaultendpunct}{\mcitedefaultseppunct}\relax
\EndOfBibitem
\bibitem[Gaggioli \latin{et~al.}(2024)Gaggioli, Blatter, Novoselov, and Geshkenbein]{gaggioli2024superconductivity}
Gaggioli,~F.; Blatter,~G.; Novoselov,~K.~S.; Geshkenbein,~V.~B. Superconductivity in atomically thin films: Two-dimensional critical state model. \emph{Physical Review Research} \textbf{2024}, \emph{6}, 023190\relax
\mciteBstWouldAddEndPuncttrue
\mciteSetBstMidEndSepPunct{\mcitedefaultmidpunct}
{\mcitedefaultendpunct}{\mcitedefaultseppunct}\relax
\EndOfBibitem
\bibitem[J{\"o}nsson(2022)]{jonsson2022theory}
J{\"o}nsson,~M. Theory for superconducting few-photon detectors. Ph.D.\ thesis, KTH Royal Institute of Technology, 2022\relax
\mciteBstWouldAddEndPuncttrue
\mciteSetBstMidEndSepPunct{\mcitedefaultmidpunct}
{\mcitedefaultendpunct}{\mcitedefaultseppunct}\relax
\EndOfBibitem
\bibitem[Mironov \latin{et~al.}(2018)Mironov, Silevitch, Proslier, Postolova, Burdastyh, Gutakovskii, Rosenbaum, Vinokur, and Baturina]{mironov2018charge}
Mironov,~A.~Y.; Silevitch,~D.~M.; Proslier,~T.; Postolova,~S.~V.; Burdastyh,~M.~V.; Gutakovskii,~A.~K.; Rosenbaum,~T.~F.; Vinokur,~V.~V.; Baturina,~T.~I. Charge Berezinskii-Kosterlitz-Thouless transition in superconducting nbtin films. \emph{Scientific reports} \textbf{2018}, \emph{8}, 4082\relax
\mciteBstWouldAddEndPuncttrue
\mciteSetBstMidEndSepPunct{\mcitedefaultmidpunct}
{\mcitedefaultendpunct}{\mcitedefaultseppunct}\relax
\EndOfBibitem
\bibitem[Zotova and Vodolazov(2014)Zotova, and Vodolazov]{zotova_intrinsic_2014}
Zotova,~A.~N.; Vodolazov,~D.~Y. Intrinsic detection efficiency of superconducting nanowire single photon detector in the modified hot spot model. \emph{Superconductor Science and Technology} \textbf{2014}, \emph{27}, 125001, Publisher: {IOP} Publishing\relax
\mciteBstWouldAddEndPuncttrue
\mciteSetBstMidEndSepPunct{\mcitedefaultmidpunct}
{\mcitedefaultendpunct}{\mcitedefaultseppunct}\relax
\EndOfBibitem
\bibitem[Semenov \latin{et~al.}(2001)Semenov, Gol’tsman, and Korneev]{semenov2001quantum}
Semenov,~A.~D.; Gol’tsman,~G.~N.; Korneev,~A.~A. Quantum detection by current carrying superconducting film. \emph{Physica C: Superconductivity} \textbf{2001}, \emph{351}, 349--356\relax
\mciteBstWouldAddEndPuncttrue
\mciteSetBstMidEndSepPunct{\mcitedefaultmidpunct}
{\mcitedefaultendpunct}{\mcitedefaultseppunct}\relax
\EndOfBibitem
\bibitem[Natarajan \latin{et~al.}(2012)Natarajan, Tanner, and Hadfield]{natarajan2012superconducting}
Natarajan,~C.~M.; Tanner,~M.~G.; Hadfield,~R.~H. Superconducting nanowire single-photon detectors: physics and applications. \emph{Superconductor science and technology} \textbf{2012}, \emph{25}, 063001\relax
\mciteBstWouldAddEndPuncttrue
\mciteSetBstMidEndSepPunct{\mcitedefaultmidpunct}
{\mcitedefaultendpunct}{\mcitedefaultseppunct}\relax
\EndOfBibitem
\bibitem[Bulaevskii \latin{et~al.}(2012)Bulaevskii, Graf, and Kogan]{bulaevskii_vortex-assisted_2012}
Bulaevskii,~L.~N.; Graf,~M.~J.; Kogan,~V.~G. Vortex-assisted photon counts and their magnetic field dependence in single-photon superconducting detectors. \emph{Physical Review B} \textbf{2012}, \emph{85}, 014505\relax
\mciteBstWouldAddEndPuncttrue
\mciteSetBstMidEndSepPunct{\mcitedefaultmidpunct}
{\mcitedefaultendpunct}{\mcitedefaultseppunct}\relax
\EndOfBibitem
\bibitem[Zhang \latin{et~al.}(2014)Zhang, You, Liu, Zhang, Zhang, Liu, Wu, He, Lv, Wang, and Xie]{zhang_characterization_2014}
Zhang,~L.; You,~L.; Liu,~D.; Zhang,~W.; Zhang,~L.; Liu,~X.; Wu,~J.; He,~Y.; Lv,~C.; Wang,~Z.; Xie,~X. Characterization of superconducting nanowire single-photon detector with artificial constrictions. \emph{{AIP} Advances} \textbf{2014}, \emph{4}, 067114\relax
\mciteBstWouldAddEndPuncttrue
\mciteSetBstMidEndSepPunct{\mcitedefaultmidpunct}
{\mcitedefaultendpunct}{\mcitedefaultseppunct}\relax
\EndOfBibitem
\bibitem[Vodolazov(2017)]{vodolazov_single-photon_2017}
Vodolazov,~D. Single-Photon Detection by a Dirty Current-Carrying Superconducting Strip Based on the Kinetic-Equation Approach. \emph{Physical Review Applied} \textbf{2017}, \emph{7}, 034014\relax
\mciteBstWouldAddEndPuncttrue
\mciteSetBstMidEndSepPunct{\mcitedefaultmidpunct}
{\mcitedefaultendpunct}{\mcitedefaultseppunct}\relax
\EndOfBibitem
\bibitem[Korneev \latin{et~al.}(2015)Korneev, Korneeva, Mikhailov, Pershin, Semenov, Vodolazov, Divochiy, Vakhtomin, Smirnov, Sivakov, Devizenko, and Goltsman]{korneev_characterization_2015}
Korneev,~A.~A.; Korneeva,~Y.~P.; Mikhailov,~M.~Y.; Pershin,~Y.~P.; Semenov,~A.~V.; Vodolazov,~D.~Y.; Divochiy,~A.~V.; Vakhtomin,~Y.~B.; Smirnov,~K.~V.; Sivakov,~A.~G.; Devizenko,~A.~Y.; Goltsman,~G.~N. Characterization of {MoSi} Superconducting Single-Photon Detectors in the Magnetic Field. \emph{{IEEE} Transactions on Applied Superconductivity} \textbf{2015}, \emph{25}, 1--4\relax
\mciteBstWouldAddEndPuncttrue
\mciteSetBstMidEndSepPunct{\mcitedefaultmidpunct}
{\mcitedefaultendpunct}{\mcitedefaultseppunct}\relax
\EndOfBibitem
\bibitem[Renema \latin{et~al.}(2015)Renema, Rengelink, Komen, Wang, Gaudio, op~'t Hoog, Zhou, Sahin, Fiore, Kes, Aarts, van Exter, de~Dood, and Driessen]{renema_effect_2015}
Renema,~J.~J.; Rengelink,~R.~J.; Komen,~I.; Wang,~Q.; Gaudio,~R.; op~'t Hoog,~K. P.~M.; Zhou,~Z.; Sahin,~D.; Fiore,~A.; Kes,~P.; Aarts,~J.; van Exter,~M.~P.; de~Dood,~M. J.~A.; Driessen,~E. F.~C. The effect of magnetic field on the intrinsic detection efficiency of superconducting single-photon detectors. \emph{Applied Physics Letters} \textbf{2015}, \emph{106}, 092602\relax
\mciteBstWouldAddEndPuncttrue
\mciteSetBstMidEndSepPunct{\mcitedefaultmidpunct}
{\mcitedefaultendpunct}{\mcitedefaultseppunct}\relax
\EndOfBibitem
\bibitem[Jahani \latin{et~al.}(2020)Jahani, Yang, Buganza~Tepole, Bardin, Tang, and Jacob]{jahani_probabilistic_2020}
Jahani,~S.; Yang,~L.-P.; Buganza~Tepole,~A.; Bardin,~J.~C.; Tang,~H.~X.; Jacob,~Z. Probabilistic vortex crossing criterion for superconducting nanowire single-photon detectors. \emph{Journal of Applied Physics} \textbf{2020}, \emph{127}, 143101\relax
\mciteBstWouldAddEndPuncttrue
\mciteSetBstMidEndSepPunct{\mcitedefaultmidpunct}
{\mcitedefaultendpunct}{\mcitedefaultseppunct}\relax
\EndOfBibitem
\bibitem[Hadfield \latin{et~al.}(2007)Hadfield, Dalgarno, O’Connor, Ramsay, Warburton, Gansen, Baek, Stevens, Mirin, and Nam]{hadfield2007submicrometer}
Hadfield,~R.~H.; Dalgarno,~P.~A.; O’Connor,~J.~A.; Ramsay,~E.; Warburton,~R.~J.; Gansen,~E.~J.; Baek,~B.; Stevens,~M.~J.; Mirin,~R.~P.; Nam,~S.~W. Submicrometer photoresponse mapping of nanowire superconducting single-photon detectors. \emph{Applied Physics Letters} \textbf{2007}, \emph{91}\relax
\mciteBstWouldAddEndPuncttrue
\mciteSetBstMidEndSepPunct{\mcitedefaultmidpunct}
{\mcitedefaultendpunct}{\mcitedefaultseppunct}\relax
\EndOfBibitem
\bibitem[Renema \latin{et~al.}(2015)Renema, Wang, Gaudio, Komen, Hoog, Sahin, Schilling, Exter, Fiore, Engel, and Dood]{renema_position-dependent_2015}
Renema,~J.~J.; Wang,~Q.; Gaudio,~R.; Komen,~I.; Hoog,~K. o.~t.; Sahin,~D.; Schilling,~A.; Exter,~M. P.~v.; Fiore,~A.; Engel,~A.; Dood,~M. J. A.~d. Position-Dependent Local Detection Efficiency in a Nanowire Superconducting Single-Photon Detector. \emph{Nano Letters} \textbf{2015}, \emph{15}, 4541--4545\relax
\mciteBstWouldAddEndPuncttrue
\mciteSetBstMidEndSepPunct{\mcitedefaultmidpunct}
{\mcitedefaultendpunct}{\mcitedefaultseppunct}\relax
\EndOfBibitem
\bibitem[Ceccarelli \latin{et~al.}(2019)Ceccarelli, Vasyukov, Wyss, Romagnoli, Rossi, Moser, and Poggio]{ceccarelli2019imaging}
Ceccarelli,~L.; Vasyukov,~D.; Wyss,~M.; Romagnoli,~G.; Rossi,~N.; Moser,~L.; Poggio,~M. Imaging pinning and expulsion of individual superconducting vortices in amorphous MoSi thin films. \emph{arXiv preprint arXiv:1907.05110} \textbf{2019}, \relax
\mciteBstWouldAddEndPunctfalse
\mciteSetBstMidEndSepPunct{\mcitedefaultmidpunct}
{}{\mcitedefaultseppunct}\relax
\EndOfBibitem
\bibitem[Xie \latin{et~al.}(2025)Xie, Fukumori, Li, and Faraon]{xie2025scalable}
Xie,~T.; Fukumori,~R.; Li,~J.; Faraon,~A. Scalable microwave-to-optical transducers at the single-photon level with spins. \emph{Nature Physics} \textbf{2025}, \emph{21}, 931--937\relax
\mciteBstWouldAddEndPuncttrue
\mciteSetBstMidEndSepPunct{\mcitedefaultmidpunct}
{\mcitedefaultendpunct}{\mcitedefaultseppunct}\relax
\EndOfBibitem
\bibitem[Tokuhara \latin{et~al.}(1985)Tokuhara, Ohtsu, Ono, Yamada, Sagawa, and Matsuura]{tokuhara1985magnetization}
Tokuhara,~K.; Ohtsu,~Y.; Ono,~F.; Yamada,~O.; Sagawa,~M.; Matsuura,~Y. Magnetization and torque measurements on Nd2Fe14B single crystals. \emph{Solid state communications} \textbf{1985}, \emph{56}, 333--336\relax
\mciteBstWouldAddEndPuncttrue
\mciteSetBstMidEndSepPunct{\mcitedefaultmidpunct}
{\mcitedefaultendpunct}{\mcitedefaultseppunct}\relax
\EndOfBibitem
\bibitem[Givord \latin{et~al.}(1984)Givord, Li, and De~La~B{\^a}thie]{givord1984magnetic}
Givord,~D.; Li,~H.; De~La~B{\^a}thie,~R.~P. Magnetic properties of Y2Fe14B and Nd2Fe14B single crystals. \emph{Solid state communications} \textbf{1984}, \emph{51}, 857--860\relax
\mciteBstWouldAddEndPuncttrue
\mciteSetBstMidEndSepPunct{\mcitedefaultmidpunct}
{\mcitedefaultendpunct}{\mcitedefaultseppunct}\relax
\EndOfBibitem
\end{mcitethebibliography}
\end{document}